\documentclass[a4paper,10pt]{article}
\usepackage[T1]{fontenc}
\usepackage{lmodern}
\usepackage{authblk}
\usepackage{amsmath, amssymb, mathrsfs, bm, amsthm, amsfonts}
\usepackage{appendix}
\usepackage{graphicx}
\usepackage{epstopdf}
\usepackage[numbers,sort&compress]{natbib}
\usepackage{tikz,pgfmath,pgfplots,subfigure}
\usetikzlibrary{math,decorations.pathreplacing,decorations.fractals,spy}
\pgfplotsset{compat=1.18}
\usepackage{booktabs}
\usepackage{multirow}
\usepackage{float}
\usepackage{geometry}
\usepackage{upgreek}
\usepackage{indentfirst}
\usepackage{setspace}
\usepackage[inline]{enumitem}
\usepackage[hidelinks,colorlinks=false,linkcolor=gray,anchorcolor=blue,citecolor=gray,bookmarks=true,bookmarksopen=true,bookmarksnumbered=true]{hyperref}%
\makeatletter
\AtBeginDocument{
  \hypersetup{
    pdftitle = {\@title},
    pdfauthor = {Fuhao Liu},
    pdfsubject={Metasurface},
    pdfkeywords={Metasurface; NtD map; Finite element method; Helmholtz's equation}
  }
}
\makeatother
\allowdisplaybreaks[4]

\usepackage{xparse}
\newcommand{\lsum}[2]{\sum\limits_{#1}^{#2}}
\newcommand{\NTC}{p}
\newcommand{\NC}{p_{\text{0}}}
\newcommand{\ttot}{\text{tot}}
\newcommand{\upref}{\text{ref}}
\newcommand{\uref}[0]{u^{\upref}}
\newcommand{\wavel}[0]{k_0}
\newcommand{\micron}[0]{\upmu\text{m}}
\NewDocumentCommand{\domain}{gg}{\Omega\IfNoValueF{#1}{_{#1}}\IfNoValueF{#2}{^{(#2)}}}
\NewDocumentCommand{\NtD}{mg}{\Lambda^{(#1)}\IfNoValueF{#2}{_{#2}}}
\NewDocumentCommand{\NtDM}{mg}{\tilde{\Lambda}^{(#1)}\IfNoValueF{#2}{_{#2}}}
\NewDocumentCommand{\RTM}{mg}{f^{(#1)}\IfNoValueF{#2}{_{#2}}}
\NewDocumentCommand{\RTMh}{mg}{\tilde{f}^{(#1)}\IfNoValueF{#2}{_{#2}}}
\NewDocumentCommand{\cmdu}{mg}{u_{#1}\IfNoValueF{#2}{^{(#2)}}}
\NewDocumentCommand{\cmdw}{mg}{w_{#1}\IfNoValueF{#2}{^{(#2)}}}
\NewDocumentCommand{\cmdpxu}{gg}{\partial_{x}u\IfNoValueF{#1}{_{#1}}\IfNoValueF{#2}{^{(#2)}}}
\newcommand{\cmdpyw}[2]{\partial_{y}w_{#1}^{(#2)}} 
\NewDocumentCommand{\cmdui}{gg}{u
  \IfNoValueT{#1}{^{\upref}}
  \IfNoValueF{#1}{_{#1}}\IfNoValueF{#2}{^{(#2)}}}
\NewDocumentCommand{\cmdpxui}{gg}{\partial_{x}u
\IfNoValueT{#1}{^{\upref}}
\IfNoValueF{#1}{_{#1}}^{(\upref\IfNoValueF{#2}{,#2})}}
\newcommand{\cmdwi}[2]{w_{#1}^{(\upref,#2)}} 
\newcommand{\cmdpywi}[2]{\partial_{y}w_{#1}^{(\upref,#2)}}

\NewDocumentCommand{\cmduh}{mgg}{\tilde{u}_{#1\IfNoValueF{#3}{,#3}}\IfNoValueF{#2}{^{(#2)}}}
\NewDocumentCommand{\cmdpxuh}{mgg}{\partial_{x}\tilde{u}_{#1\IfNoValueF{#3}{,#3}}\IfNoValueF{#2}{^{(#2)}}}
\NewDocumentCommand{\cmdwh}{mgg}{\tilde{w}_{#1\IfNoValueF{#3}{,#3}}^{(#2)}}
\NewDocumentCommand{\cmdpywh}{mgg}{\partial_{y}\tilde{w}_{#1\IfNoValueF{#3}{,#3}}^{(#2)}}
\NewDocumentCommand{\cmduih}{mgg}{\tilde{u}_{#1\IfNoValueF{#3}{,#3}}^{(\upref\IfNoValueF{#2}{,#2})}}
\NewDocumentCommand{\cmdpxuih}{mgg}{\partial_{x}\tilde{u}_{#1\IfNoValueF{#3}{,#3}}^{(\upref\IfNoValueF{#2}{,#2})}}
\NewDocumentCommand{\cmdwih}{mgg}{\tilde{w}_{#1\IfNoValueF{#3}{,#3}}^{(\upref\IfNoValueF{#2}{,#2})}}
\NewDocumentCommand{\cmdpywih}{mgg}{\partial_{y}\tilde{w}_{#1\IfNoValueF{#3}{,#3}}^{(\upref\IfNoValueF{#2}{,#2})}}
\NewDocumentCommand{\ulist}{gg}{
  \IfNoValueTF{#1}{v}{\tilde{v}}
  \IfNoValueF{#2}{_{#2}}\IfNoValueF{#1}{^{(\text{#1})}}}

\title{An efficient numerical method for simulating two-dimensional non-periodic metasurfaces }

\author[1]{
  Fuhao Liu \thanks{Corresponding author: $\mathtt{fuhaoliu@cityu.edu.hk}$}
}
\author[1,2]{
  Ya Yan Lu 
}
\affil[1]{\small Liu Bie Ju Centre for Mathematical Sciences, City University of Hong Kong, Kowloon, Hong Kong, China}
\affil[2]{\small Department of Mathematics, City University of Hong Kong, Kowloon, Hong Kong, China}

\begin{document}
\maketitle

\begin{abstract}
Metasurfaces are extremely useful for controlling and manipulating electromagnetic waves. 
Full-wave numerical simulation is highly desired for their design and optimization, but it is notoriously difficult, even for two-dimensional metasurfaces, when they comprise a huge number of subwavelength elements. This paper focuses on two-dimensional non-periodic metasurfaces that contain only a relatively small number of distinct subwavelength elements. We develop an efficient numerical method based on  Neumann-to-Dirichlet operators, the finite element method and local function expansions. 
Our method drastically reduces the total number of unknowns and is capable of simulating two-dimensional metasurfaces with $10^{5}$ subwavelength elements on a personal computer. Numerical examples demonstrate that the method maintains high accuracy while offering significant advantages in both computational time and memory usage compared to the classical full-domain finite element method, making it particularly suited for the analysis of large metasurfaces.

  \vspace{10pt}
\end{abstract}

\section{Introduction}

Metasurfaces are structures obtained by arranging a huge number of subwavelength dielectric or metallic elements (often called resonators or scatterers) on a substrate or slab \cite{2013-Kildishev-S,
2017-Khorasaninejad-S,
2025-Brongersma-NREE}.
They provide exceptional control over electromagnetic waves, allowing precise manipulation of their phase, amplitude, polarization, and wavefront. Due to their compact form and relatively simple fabrication process, metasurfaces have found numerous applications, such as metalenses, beam deflectors, splitters, invisible cloaks, holograms, waveplates, optical mode converters \cite{
  2014-Yu-NM,
  2015-Ni-S,
  2016-Khorasaninejad-S,
  2017-Khorasaninejad-S,
  2022-Santiago-Cruz-S,
  2025-Brongersma-NREE,
  2025-Soma-NC,
  2025-Gu-N}.

The design of metasurfaces typically follows a multi-step process. Initially, a set of subwavelength elements is designed or selected. These elements are then arranged on a substrate or slab in a predefined pattern. Subsequently, the entire device is numerically simulated to verify its performance. If the metasurface fails to meet the target specifications, the arrangement and simulation steps are iterated as necessary. Numerical simulation plays a critical role in metasurface development, since it is essential for verifying performance and mitigating high manufacturing costs.
In principle, full-wave simulation should be used for modeling metasurfaces. For relatively small metasurfaces, several classical numerical methods have been employed. Notable examples include the finite-difference time-domain (FDTD) \cite{2017-Ding-Wave,2025-Huang-JCAM} and discontinuous Galerkin time-domain (DGTD) \cite{2023-Chen-IToAaP,2025-Li-JoCP} methods, along with the finite-difference method (FDM) \cite{2016-Vahabzadeh-ITAP}, finite element method (FEM) \cite{2017-Sandeep-ITAP,2023-Yang-CPC,2023-Amboli-OE,2026-Mao-JoCaAM}, integral equation method \cite{2015-Francavilla-IToAaP} and spectral element method (SEM) \cite{2023-Cai-IToAaP} in the frequency domain.
However, the size of a practical metasurface is usually very large (on the order of $10^4$ wavelengths or more), hence these classical methods often require prohibitively large computer memory and extremely long computation time. This makes full-wave simulation of metasurfaces a very challenging problem. To tackle this difficulty, some improved or alternative methods are proposed, such as domain decomposition method \cite{2009-Dolean-SJoSC,2013-Chen-SJoNA,2016-Tao-MC,2013-Stolk-JoCP,2019-Gander-SR,2019-Leng-SJoSC,2024-Gao-ITAP}, numerical mode-matching method \cite{2019-Liu-IToMTaT,2020-Liu-ITMTT,2024-Zhang-PRE}, {\color{red}axisymmetric FEM \cite{2020-Christiansen-OE}}, integral equations method with fast direct solvers \cite{2023-Xue-apa} and parallel-accelerated FDTD method \cite{2010-Oskooi-CPC,2019-Warren-CPC,2021-Hughes-APL}. Notably, this last method underpins the GPU-accelerated engines in many commercial solvers, such as Ansys Lumerical FDTD, XFdtd, and Tidy3D.
Some of these methods are primarily suited for problems with special structures, while others may require many iterations. 
To avoid the difficulty of full-wave simulation, some researchers developed local methods, such as the local periodic approximation method \cite{2016-Byrnes-OE,2018-Chen-NN,2018-Pestourie-OE} and the overlapping domain approximation approach \cite{2019-Lin-OE}. However, these local methods always give rise to relatively large simulation errors.

In this paper, we study two-dimensional (2D) metasurfaces which are translationally invariant in one spatial direction. Full-wave numerical simulation of a large 2D metasurface is still very difficult, when the size of the metasurface is $10^4$ wavelengths or more. We focus on non-periodic 2D metasurfaces with a very large number of unit cells (each containing a subwavelength element), assuming that the number of distinct unit cells is relatively small. Notice that a metasurface is an engineered structure based on pre-designed subwavelength elements. The number of distinct subwavelength elements is usually relatively small. Taking advantage of this structural character, we develop an efficient full-wave numerical method based on the concept of Neumann-to-Dirichlet (NtD) operator.
For a given domain $\domain$ and a linear homogeneous differential equation, the NtD operator maps the normal derivative of any solution $u$ to $u$ itself on the boundary of $\domain$. In numerical implementations, the NtD operator is approximated by a matrix. We use FEM to calculate the NtD matrices. To reduce the size of the NtD matrices, we further use local function expansions on the boundary of the domain. In previous works, NtD- or DtN-based methods have been used to solve scattering problems \cite{2008-Hu-OE} and calculate waveguide modes \cite{2012-Lu-JCP}.
A 2D metasurface, although very large in one spatial direction, is still finite in the 2D plane. In our implementation, we surround the metasurface with perfectly matched layers (PMLs), divide the truncated structure into unit cells, and derive a linear system for the normal derivative of the field on the interfaces between the unit cells.
Once the linear system is solved, we can easily reconstruct the wave field within every unit cell. Numerical examples confirm that our method is accurate and efficient for simulating large 2D metasurfaces.

The structure of this paper is as follows. Section 2 provides an overview of our method, including the definition of the NtD operators and the derivation of the final linear system. Section 3 details the calculation of a relationship between the fields at two edges of each unit cell and the incident wave.  In Sec. 4, local function expansions are combined with FEM to obtain small matrix approximations of the NtD operators. Section 5 describes the procedure for recovering the far field from the near field. In Sec. 6, the accuracy and efficiency of our method are verified by simulating a gradient-phase metasurface, with results compared to those from the classical FEM. We also present a comparison of the computation time and peak memory consumption required by both methods. Finally, Section 7 concludes the paper.

\section{Overview of the method}
Consider a non-periodic 2D metasurface, which consists of tens of thousands of non-magnetic objects on an infinite dielectric slab in free space.
For 2D structures that are invariant in the $z$-direction, if waves propagate
in the $x$-$y$ plane, the governing equation for the $z$-component of the
electric field $u$ is the Helmholtz equation
\begin{eqnarray}\label{Helmholtz:eqn}
  \nabla\cdot(\nabla u)+\wavel^{2} n^2 u=0,
\end{eqnarray}
where $\wavel$ is the wavenumber in free space, $n=n(x,y)$ is the refractive index function, and $\nabla=(\partial_{x},\partial_{y})^{\top}$ is the gradient operator.
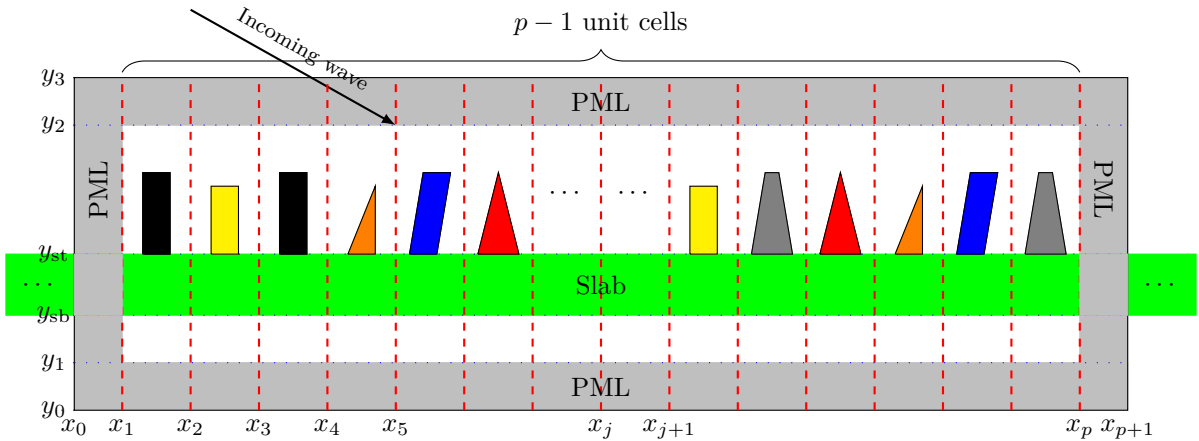
\begin{figure}
  \centering
  \begin{tikzpicture}[scale=0.9]
    \def\height{{1,1.2}}
    \def\yp{{-0.7,0,0.7,1.6,3.5,4.2}}
    \def\xp{{-0.7,0,1,2,3,4,5,6,7,8,9,10,11,12,13,14,14.7}}
    \tikzmath{ \xpm=-1.0; }
    \tikzmath{ \xpN=16; }
    \filldraw[color=lightgray] (\xp[0],\yp[0])  rectangle (\xp[16],\yp[1]);
    \filldraw[color=lightgray] (\xp[0],\yp[4]) rectangle (\xp[16],\yp[5]);
    \filldraw[color=lightgray] (\xp[0],\yp[0]) rectangle (\xp[1],\yp[5]);
    \filldraw[color=lightgray] (\xp[15],\yp[0]) rectangle (\xp[16],\yp[5]);
    \draw[black,-latex,thick] (\xp[2],5.2) -- (\xp[5],\yp[4]) node[midway, sloped, above] {\footnotesize Incoming wave};
    \filldraw[green] (\xp[0]-1,\yp[2]) rectangle (\xp[\xpN]+1,\yp[3]);
    \draw[black,thin] (\xp[0],\yp[0]) rectangle (\xp[\xpN],\yp[5]);

    \foreach \i in {1,...,{\numexpr \xpN-1\relax}}      {
        \draw[red,thick,dashed] (\xp[\i],\yp[0]) -- (\xp[\i],\yp[5]);
      }
    \foreach \i in {1,...,4}{
        \draw[blue,thin,loosely dotted] (\xp[0],\yp[\i]) -- (\xp[\xpN],\yp[\i]);
      }
    \foreach \i in {4,12}{
        \coordinate (P1) at (\xp[\i]*0.7+\xp[\i+1]*0.3,\yp[3]);
        \coordinate (P2) at (\xp[\i]*0.3+\xp[\i+1]*0.7,\yp[3]);
        \coordinate (P3) at (\xp[\i]*0.3+\xp[\i+1]*0.7,\yp[3]+\height[0]);
        \filldraw[orange] (P1) -- (P2) -- (P3) -- (P1);
        \draw (P1) -- (P2) -- (P3) -- (P1);
      }
    \foreach \i in {2,9}{
        \coordinate (P1) at (\xp[\i]*0.7+\xp[\i+1]*0.3,\yp[3]);
        \coordinate (P2) at (\xp[\i]*0.3+\xp[\i+1]*0.7,\yp[3]);
        \coordinate (P3) at (\xp[\i]*0.3+\xp[\i+1]*0.7,\yp[3]+\height[0]);
        \coordinate (P4) at (\xp[\i]*0.7+\xp[\i+1]*0.3,\yp[3]+\height[0]);
        \filldraw[yellow] (P1) -- (P2) -- (P3) -- (P4) -- (P1);
        \draw (P1) -- (P2) -- (P3) -- (P4) -- (P1);
      }
    \foreach \i in {1,3}{
        \coordinate (P1) at (\xp[\i]*0.7+\xp[\i+1]*0.3,\yp[3]);
        \coordinate (P2) at (\xp[\i]*0.3+\xp[\i+1]*0.7,\yp[3]);
        \coordinate (P3) at (\xp[\i]*0.3+\xp[\i+1]*0.7,\yp[3]+\height[1]);
        \coordinate (P4) at (\xp[\i]*0.7+\xp[\i+1]*0.3,\yp[3]+\height[1]);
        \filldraw[black] (P1) -- (P2) -- (P3) -- (P4) -- (P1);
        \draw (P1) -- (P2) -- (P3) -- (P4) -- (P1);
      }
    \foreach \i in {10,14}{
        \coordinate (P1) at (\xp[\i]*0.8+\xp[\i+1]*0.2,\yp[3]);
        \coordinate (P2) at (\xp[\i]*0.2+\xp[\i+1]*0.8,\yp[3]);
        \coordinate (P3) at (\xp[\i]*0.4+\xp[\i+1]*0.6,\yp[3]+\height[1]);
        \coordinate (P4) at (\xp[\i]*0.6+\xp[\i+1]*0.4,\yp[3]+\height[1]);
        \filldraw[gray] (P1) -- (P2) -- (P3) -- (P4) -- (P1);
        \draw (P1) -- (P2) -- (P3) -- (P4) -- (P1);
      }
    \foreach \i in {6,11}{
        \coordinate (P1) at (\xp[\i]*0.8+\xp[\i+1]*0.2,\yp[3]);
        \coordinate (P2) at (\xp[\i]*0.2+\xp[\i+1]*0.8,\yp[3]);
        \coordinate (P3) at (\xp[\i]*0.5+\xp[\i+1]*0.5,\yp[3]+\height[1]);
        \filldraw[red] (P1) -- (P2) -- (P3) -- (P1);
        \draw (P1) -- (P2) -- (P3) -- (P1);
      }
    \foreach \i in {5,13}{
        \coordinate (P1) at (\xp[\i]*0.8+\xp[\i+1]*0.2,\yp[3]);
        \coordinate (P2) at (\xp[\i]*0.4+\xp[\i+1]*0.6,\yp[3]);
        \coordinate (P3) at (\xp[\i]*0.2+\xp[\i+1]*0.8,\yp[3]+\height[1]);
        \coordinate (P4) at (\xp[\i]*0.6+\xp[\i+1]*0.4,\yp[3]+\height[1]);
        \filldraw[blue] (P1) -- (P2) -- (P3) -- (P4) -- (P1);
        \draw (P1) -- (P2) -- (P3) -- (P4) -- (P1);
        
      }
    \node at (\xp[7]/2+\xp[8]/2,2.5) {$\cdots$};
    \node at (\xp[8]/2+\xp[9]/2,2.5) {$\cdots$};
    \node at (\xp[0]-0.5,\yp[2]/2+\yp[3]/2) {$\cdots$};
    \node at (\xp[\xpN]+0.5,\yp[2]/2+\yp[3]/2) {$\cdots$};
    \node at (\xp[1]/2+\xp[\xpN-1]/2,\yp[2]/2+\yp[3]/2) {Slab};
    \node at (\xp[1]/2+\xp[\xpN-1]/2,\yp[0]/2+\yp[1]/2) {PML};
    \node at (\xp[1]/2+\xp[\xpN-1]/2,\yp[4]/2+\yp[5]/2) {PML};
    \node[rotate=90] at (\xp[0]/2+\xp[1]/2,\yp[3]/2+\yp[4]/2) {PML};
    \node[rotate=-90] at (\xp[\xpN-1]/2+\xp[\xpN]/2,\yp[3]/2+\yp[4]/2) {PML};
    \draw[loosely dotted] (\xp[0],\yp[0]) -- (\xpm,\yp[0])  node {$y_0$};
    \draw[loosely dotted] (\xp[0],\yp[1]) -- (\xpm,\yp[1])  node {$y_1$};
    \draw[loosely dotted] (\xp[0],\yp[2]) -- (\xpm,\yp[2])  node {$y_{\text{sb}}$};
    \draw[loosely dotted] (\xp[0],\yp[3]) -- (\xpm,\yp[3])  node {$y_{\text{st}}$};
    \draw[loosely dotted] (\xp[0],\yp[4]) -- (\xpm,\yp[4])  node {$y_2$};
    \draw[loosely dotted] (\xp[0],\yp[5]) -- (\xpm,\yp[5])  node {$y_3$};
    \foreach \i in {0,...,5}{
        \node[below] at (\xp[\i],\yp[0]) {$x_{\i}$};
      }
    \node[below] at (\xp[8],\yp[0]) {$x_{j}$};
    \node[below] at (\xp[9],\yp[0]) {$x_{j+1}$};
    \node[below] at (\xp[\xpN-1],\yp[0]) {$x_{\NTC}$};
    \node[below] at (\xp[\xpN],\yp[0]) {$x_{\NTC+1}$};
\filldraw[color=lightgray] (\xp[0],\yp[2]) rectangle (\xp[1],\yp[3]) ;
\filldraw[color=lightgray] (\xp[15],\yp[2]) rectangle (\xp[16],\yp[3]) ;
    \draw[decorate,decoration={brace,raise=2pt,amplitude=0.3cm}] (\xp[1],\yp[5]) -- (\xp[\xpN-1],\yp[5]) node at (\xp[1]/2+\xp[\xpN-1]/2,\yp[5]+0.8) {$\NTC-1$ unit cells};

  \end{tikzpicture}
  \caption{Sketch of a non-periodic 2D metasurface, which can be divided into $\NTC-1$ unit cells.}\label{Sketch:Metasurface}
\end{figure}

As shown in Fig. \ref{Sketch:Metasurface}, in the $y$ direction, the
metasurface is contained in the interval $(y_1,y_2)$ where the medium for
$y<y_1$ and $y>y_2$ is free space. We assume that the metasurface can be
divided into $\NTC-1$ unit cells, where the $j$-th unit cell corresponds to
$x\in(x_{j},x_{j+1})$. For $x<x_{1}$ and $x>x_{\NTC}$, the structure is a
simple uniform slab with the refractive index $n$ depending on $y$ only, that is
\begin{eqnarray*}
  n(y)=\begin{cases}
    n_{\text{s}}>1, & y_{\text{sb}}<y<y_{\text{st}}, \\
    1,       & \text{otherwise},
  \end{cases}
\end{eqnarray*}
where $y_{1}\leq y_{\text{sb}}<y_{\text{st}}<y_{2}$.
Therefore, the metasurface is contained in the rectangular domain $(x_{1},x_{\NTC})\times(y_1,y_2)$, where $x_{\NTC}-x_{1}$ is significantly larger than $y_2-y_1$. We consider a scattering problem, where the metasurface is illuminated from above by an incoming wave. The incoming wave is a plane wave given by
\begin{eqnarray}\label{u_inc}
  u^{\text{inc}}(x,y)=\exp(i(\alpha x-\beta y)), &\alpha=\wavel \sin(\theta), ~~\beta=\wavel \cos(\theta),
\end{eqnarray}
where $i=\sqrt{-1}$, and $\theta$ is the incident angle.

To truncate the infinite media, we use a perfectly matched layer (PML) to
surround the rectangular domain $(x_1,x_{\NTC})\times(y_1,y_2)$. The PML
corresponds to the intervals $(x_0,x_1)$, $(x_{\NTC},x_{\NTC+1})$, $(y_0,y_1)$
and $(y_{2},y_{3})$. Therefore, the entire computational domain is the rectangle
$(x_{0},x_{\NTC+1})\times(y_0,y_3)$. For the incident wave given in Eq.
\eqref{u_inc}, if there are no objects on the slab, we can easily find the
reflected and transmitted waves. The exact solution $\uref$ is:
\begin{eqnarray}\label{Solution:ref}
  \uref =\begin{cases}
    \exp(i[\alpha x-\beta (y-y_{\text{st}})])+C_0\exp(i[\alpha x+\beta (y-y_{\text{st}})]),        & y\geq y_{\text{st}},           \\
    C_1\exp(i[\alpha x-\beta_{\text{s}} (y-y_{\text{st}})])+C_2\exp(i[\alpha x+\beta_{\text{s}} (y-y_{\text{sb}})]), & y_{\text{sb}}<y<y_{\text{st}}, \\
    C_3\exp(i[\alpha x-\beta (y-y_{\text{sb}})]),                                                  & y\leq y_{\text{sb}},           \\
  \end{cases}
\end{eqnarray}
where $\beta_{\text{s}}=\sqrt{(\wavel n_{\text{s}})^2 - \alpha^2}$, and the coefficients $C_0, C_1, C_2, C_3$ satisfy a system of equations obtained from the continuity of $\uref$ and $\partial_y\uref$ at $y=y_{\text{sb}}$ and $y=y_{\text{st}}$.

Since the PML is used to simulate outgoing waves and the total wave $u^{\ttot}$
(containing the incident wave) is not outgoing, we need to subtract $\uref$
from the total wave in the PML region. To simplify the notation, we let
\begin{eqnarray}\label{Definition:u}
  u=\begin{cases}
    u^{\ttot}-\uref, & \text{in PML},                              \\
    u^{\ttot},       & \text{in } (x_{1},x_{\NTC})\times(y_1,y_2).
  \end{cases}
\end{eqnarray}

The entire computational domain $(x_{0},x_{\NTC+1})\times (y_0,y_3)$ can be
divided into $\NTC+1$ subdomains
\begin{eqnarray*}
  \domain{j} = \{(x,y)|x_j< x < x_{j+1},y_0 < y< y_{3}\}, ~~j=0,1,2,\cdots,\NTC,
\end{eqnarray*}
where $\Omega_{0}$ and $\Omega_{\NTC}$ are the left and right PML regions. The interfaces between the subdomains are
\begin{eqnarray*}
  \Gamma_{j}=\{(x,y)|x=x_{j}  ,y_{0} < y < y_{3}\}, ~~j=1,2,\cdots,\NTC.
\end{eqnarray*}

Our method involves two steps. In the first step, we calculate an operator
$\NtD{j}$ and a function $\RTM{j}$ for each unit cell $\domain{j}$. In the
second step, we set up a block tridiagonal linear system for $\cmdpxu$ at
all $\Gamma_{j}$. Here, $\NtD{j}$ ($1\leq j\leq\NTC-1$) is the NtD operator of
$\domain{j}$, which maps $\cmdpxu$ at $\Gamma_j$ and $\Gamma_{j+1}$ to $u$ at
$\Gamma_j$ and $\Gamma_{j+1}$. The function $\RTM{j}$ accounts for the incident
wave. The relationship between $u$ and $\cmdpxu$ at $\Gamma_j$ and $\Gamma_{j+1}$ will be
\begin{eqnarray}\label{NtD:Omegaj}
  \begin{bmatrix}
    u_{j}   \\
    u_{j+1} \\
  \end{bmatrix}=\NtD{j}\begin{bmatrix}
    \cmdpxu{j}   \\
    \cmdpxu{j+1} \\
  \end{bmatrix}+
  \RTM{j} ,~\NtD{j}=\begin{bmatrix}
    \NtD{j}{11} & \NtD{j}{12} \\
    \NtD{j}{21} & \NtD{j}{22} \\
  \end{bmatrix}, ~\RTM{j}{} =\begin{bmatrix}
    \RTM{j}{1} \\
    \RTM{j}{2} \\
  \end{bmatrix},~j=1,2,\cdots,\NTC-1,
\end{eqnarray}
where $u_j$, $u_{j+1}$, $\cmdpxu{j}$, $\cmdpxu{j+1}$ denote $u$ and $\cmdpxu$ on $\Gamma_j$ and $\Gamma_{j+1}$, respectively. For the left and right PML regions, the relationships are
\begin{eqnarray}
  \cmdu{1}  =\NtD{0}  \cmdpxu{1} + \RTM{0}{} , ~
  \cmdu{\NTC} =\NtD{\NTC} \cmdpxu{\NTC} + \RTM{\NTC}{}.\label{NtD:0andT}
\end{eqnarray}

Evaluating $\cmdu{j+1}$ ($1\leq j\leq \NTC-2$) from Eq. \eqref{NtD:Omegaj} and
a similar equation for $\Omega_{j+1}$, we obtain
\begin{eqnarray}\label{Relation:jandjp1}
  \NtD{j}{21}\cmdpxu{j}+(\NtD{j}{22}-\Lambda^{(j+1)}_{11})\cmdpxu{j+1}-\Lambda^{(j+1)}_{12}\cmdpxu{j+2} =\RTM{j+1}{1}-\RTM{j}{2},
\end{eqnarray}
Similarly, by evaluating $\cmdu{1}$ and $\cmdu{\NTC}$ from Eq. \eqref{NtD:0andT}, we obtain
\begin{eqnarray}\label{Relation:0andT}
  \begin{aligned}
    (\NtD{0}-\NtD{1}{11})\cmdpxu{1}-\NtD{1}{12}\cmdpxu{2} = \RTM{1}{1}-\RTM{0}, \\
    \NtD{\NTC-1}{21}\cmdpxu{\NTC-1}+(\NtD{\NTC-1}{22}-\NtD{\NTC})\cmdpxu{\NTC}=\RTM{\NTC}-\RTM{\NTC-1}{2}.
  \end{aligned}
\end{eqnarray}

Equations \eqref{Relation:0andT} and \eqref{Relation:jandjp1} for
$j=1,2,\cdots,\NTC-2$ can be rewritten as the following block tridiagonal
linear system
\begin{eqnarray}\label{Eqn:FinalTri}
  \begin{bmatrix}
    \Pi_1 & -\NtD{1}{12}  &                  &                    &                   \\
    \NtD{1}{21} & \Pi_{2} & -\NtD{2}{12}     &                    &                   \\
                & \ddots        & \ddots           & \ddots             &                   \\
                &               & \NtD{\NTC-2}{21} & \Pi_{\NTC-1} & -\NtD{\NTC-1}{12} \\
                &               &                  & \NtD{\NTC-1}{21}   & \Pi_{\NTC}  \\
  \end{bmatrix}\begin{bmatrix}
    \cmdpxu{1}      \\
    \cmdpxu{2}      \\ \vdots\\
    \cmdpxu{\NTC-1} \\
    \cmdpxu{\NTC}   \\
  \end{bmatrix}=
  \begin{bmatrix}
    \RTM{1}{1}-\RTM{0}              \\
    \RTM{2}{1}-\RTM{1}{2}           \\
    \vdots                          \\
    \RTM{\NTC-1}{1}-\RTM{\NTC-2}{2} \\
    \RTM{\NTC}-\RTM{\NTC-1}{2}      \\
  \end{bmatrix},
\end{eqnarray}
where $\Pi_1=\NtD{0}{}-\NtD{1}{11}$, $\Pi_{\NTC}=\Lambda^{(\NTC-1)}_{22}-\NtD{\NTC}$, $\Pi_{j}=\NtD{j-1}{22}-\NtD{j}{11}$, $j=1,2,\cdots,\NTC-1$.

Each $\NtD{j}$ in this section can be approximated by a small matrix, and each
$\RTM{j}{}$ can be approximated by a column vector. Since those $\NtD{j}$ for
identical unit cells are the same, they require computation only once and are
parallelizable. Although all $\RTM{j}{}$ are different, they can be calculated
efficiently and in parallel. In this paper, we assume that there are only $\NC$
distinct unit cells. Therefore, if $\NC\ll\NTC$, our method is efficient.

\section{Boundary relationships for unit cells}
\begin{figure}[H]
  \centering
  \subfigure[Left PML $\Omega_{0}$]{\label{Sketch:PML:left}
    \begin{tikzpicture}[scale=0.8]
      \def\yp{{-1.0,0.5,3.5,5}}
      \def\xp{{-1,1}}
      \filldraw[color=lightgray] (\xp[0],\yp[0]) rectangle (\xp[1],\yp[3]);
      \draw[black,thin] (\xp[1],\yp[0])--(\xp[0],\yp[0]) --(\xp[0],\yp[3]) -- (\xp[1],\yp[3]);
      \draw[red,thick,dashed] (\xp[1],\yp[0]) -- (\xp[1],\yp[3]);
      \foreach \i in {0,...,3}{\node[right] at (\xp[1],\yp[\i]) {$y_{\i}$};}
      \node[rotate=90] at (\xp[0]/2+\xp[1]/2,\yp[2]/2+\yp[1]/2) {PML};
      \node[below] at (\xp[0],\yp[0]) {$x_{0}$};
      \node[below] at (\xp[1],\yp[0]) {$x_{1}$};
    \end{tikzpicture}
  }
  \subfigure[Cell $\domain{j}$]{\label{Sketch:Sub:Cellj}
    \begin{tikzpicture}[scale=0.8]
      \def\xp{{1,4}}
      \def\yp{{-1.0,0.5,3.5,5}}
      \filldraw[color=lightgray] (\xp[0],\yp[0]) rectangle (\xp[1],\yp[1]);
      \filldraw[color=lightgray] (\xp[0],\yp[2]) rectangle (\xp[1],\yp[3]);
      \draw[dashed] (\xp[0],\yp[0]) -- (\xp[0],\yp[3]);
      \draw[dashed] (\xp[1],\yp[0]) -- (\xp[1],\yp[3]);
      \draw[dashed] (\xp[0],\yp[1]) -- (\xp[1],\yp[1]);
      \draw[dashed] (\xp[0],\yp[2]) -- (\xp[1],\yp[2]);
      \draw[] (\xp[0],\yp[0]) -- (\xp[1],\yp[0]);
      \draw[] (\xp[0],\yp[3]) -- (\xp[1],\yp[3]);
      \foreach \i in {0,1,2}{
          \node[] at (\xp[0]*0.4+\xp[1]*0.6,\yp[\i]/2+\yp[\i+1]/2) {$\domain{j}{\i}$};
        }
      \node[below] at (\xp[0],\yp[0]) {$x_j$};
      \node[below] at (\xp[1],\yp[0]) {$x_{j+1}$};
    \end{tikzpicture}
  }
  \subfigure[Right PML $\Omega_{\NTC}$]{\label{Sketch:PML:right}
    \begin{tikzpicture}[scale=0.8]
      \def\yp{{-1.0,0.5,3.5,5}}
      \def\xp{{10,12}}
      \filldraw[color=lightgray] (\xp[0],\yp[0]) rectangle (\xp[1],\yp[3]);
      \draw[black,thin] (\xp[0],\yp[0])--(\xp[1],\yp[0]) --(\xp[1],\yp[3]) -- (\xp[0],\yp[3]);
      \draw[red,thick,dashed] (\xp[0],\yp[0]) -- (\xp[0],\yp[3]);
      \foreach \i in {0,...,3}{\node[left] at (\xp[0],\yp[\i]) {$y_{\i}$};}
      \node[rotate=90] at (\xp[0]/2+\xp[1]/2,\yp[2]/2+\yp[1]/2) {PML};
      \node[below] at (\xp[0],\yp[0]) {$x_{\NTC}$};
      \node[below] at (\xp[1],\yp[0]) {$x_{\NTC+1}$};
    \end{tikzpicture}
  }
  \caption{The sketches of the $j$-th cell and the left and right PML.}\label{Sketch:Sub}
\end{figure}
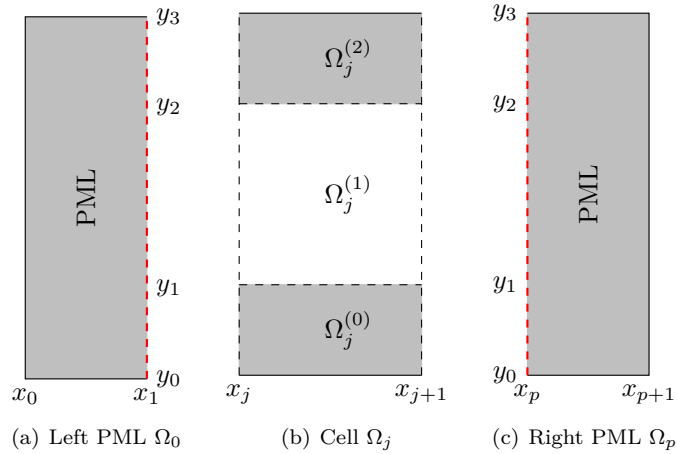

In this section, we discuss the calculation of $\NtD{j}$ and $\RTM{j}{}$ in
Eqs. \eqref{NtD:Omegaj} and \eqref{NtD:0andT}. According to Eq.
\eqref{Definition:u}, we know that $u$ is discontinuous on the interface between
the PML region and the rectangle $(x_1,x_{\NTC})\times(y_1,y_2)$. Therefore, as
shown in Fig. \ref{Sketch:Sub:Cellj}, domain $\domain{j}$ ($1\leq j\leq \NTC-1$) needs to be divided into three subdomains
\begin{eqnarray*}
  \domain{j}{k}=\{(x,y)|x_j < x<x_{j+1}, y_{k}< y<y_{k+1}\}, ~~ k=0,1,2.
\end{eqnarray*}

Since the PML transforms outgoing waves into exponentially decaying solutions, we can assume $u=0$ at the exterior boundary of the PML region. Within each subdomain $\Omega_{j}^{(k)}$, $u$ satisfies the following equation:
\begin{eqnarray}\label{Helmholtz:PML}
  \nabla\cdot(S\nabla u)+\wavel^2 n^2 s_x s_y u=0,
\end{eqnarray}
where $S$, $s_x$ and $s_y$ are
\begin{eqnarray}\label{PML:coefficient}
  S=\begin{bmatrix}
    s_y/s_x & \\& s_x/s_y\\
  \end{bmatrix}, ~~ s_x(x)=1+i\sigma_{x}(x),~~s_y(y)=1+i\sigma_{y}(y).
\end{eqnarray}
In Eq. \eqref{PML:coefficient}, $\sigma_x$ and $\sigma_y$ are real functions given as
\begin{eqnarray*}
  \sigma_x(x)=\begin{cases}
    b\dfrac{(x_1-x)^4}{(x_1-x_0)^4},                  & x_0\leq x\leq x_1,             \\
    b\dfrac{(x-x_{\NTC})^4}{(x_{\NTC+1}-x_{\NTC})^4}, & x_{\NTC}\leq x\leq x_{\NTC+1}, \\
    0,                                                         & \text{otherwise},              \\
  \end{cases}~~
  \sigma_y(y)=\begin{cases}
    b\dfrac{(y_1-y)^4}{(y_1-y_0)^4},   & y_0\leq y\leq y_1,     \\
    b\dfrac{(y-y_{2})^4}{(y_3-y_2)^4}, & y_{2}\leq y\leq y_{3}, \\
    0,                                          & \text{otherwise},      \\
  \end{cases}
\end{eqnarray*}
where $b$ denotes a positive constant
\cite{1996-Berenger-JoCP,1996-Gedney-IToAaP,2024-Jiang-AMC}.

To obtain Eq. \eqref{NtD:Omegaj}, we first calculate the NtD operator $\NtD{j,k}{}$ for each subdomain $\Omega_{j}^{(k)}$ (with $k=0,1,2$). To simplify
notation, we let
\begin{eqnarray*}
  \cmdu{j}{k}=u(x_{j},y),~\cmdpxu{j}{k}=\partial_{x}u(x_{j},y), ~ y\in(y_{k},y_{k+1}), ~~j=1,2,\cdots,\NTC, ~~~k=0,1,2.
\end{eqnarray*}
Notice that for all $j$, $\cmdu{j}{1}$ and $\cmdpxu{j}{1}$ represent the total solution that contains $\uref$. In addition, we define
\begin{eqnarray*}
  \cmdw{j}{k,\pm}=u(x,y_{k}^{\pm}), ~
  \cmdpyw{j}{k,\pm}=\partial_{y}u(x,y_{k}^{\pm}),~ \cmdwi{j}{k}=\uref(x,y_{k}),~\cmdpywi{j}{k}=\partial_{y}\uref(x,y_{k}),
\end{eqnarray*}
where $x\in(x_{j},x_{j+1}), ~~j=1,2,\cdots,\NTC-1, ~k=1,2$. Here, $u(x,y_{k}^{\pm})$ represents the limit as $u$ approaches $y=y_k$ from below or above.

To construct the NtD operator $\NtD{j,1}$ ($1\leq j\leq \NTC-1$), we need to
order the four edges of $\domain{j}{1}$. If the four edges are ordered as
``bottom, left, right, and top'', then $\NtD{j,1}{}$ satisfies
\begin{eqnarray}\label{NtD:j1}
  \NtD{j,1}{}
  \begin{bmatrix}
    \cmdpyw{j}{1,+} \\
    \cmdpxu{j}{1}   \\
    \cmdpxu{j+1}{1} \\
    \cmdpyw{j}{2,-} \\
  \end{bmatrix}=\begin{bmatrix}
    \NtD{j,1}{11} & \NtD{j,1}{12} & \NtD{j,1}{13} & \NtD{j,1}{14} \\
    \NtD{j,1}{21} & \NtD{j,1}{22} & \NtD{j,1}{23} & \NtD{j,1}{24} \\
    \NtD{j,1}{31} & \NtD{j,1}{32} & \NtD{j,1}{33} & \NtD{j,1}{34} \\
    \NtD{j,1}{41} & \NtD{j,1}{42} & \NtD{j,1}{43} & \NtD{j,1}{44} \\
  \end{bmatrix}\begin{bmatrix}
    \cmdpyw{j}{1,+} \\
    \cmdpxu{j}{1}   \\
    \cmdpxu{j+1}{1} \\
    \cmdpyw{j}{2,-} \\
  \end{bmatrix}=\begin{bmatrix}
    \cmdw{j}{1,+} \\
    \cmdu{j}{1}   \\
    \cmdu{j+1}{1} \\
    \cmdw{j}{2,-} \\
  \end{bmatrix}.
\end{eqnarray}
For $\domain{j}{0}$ and $\domain{j}{2}$, the corresponding NtD operators are defined on only three edges where $u$ is unknown. For $\NtD{j,0}$, the three edges are ordered as ``left, right, and top''. Similarly, the three edges of $\domain{j}{2}$ are ordered as ``bottom, left and right''. Therefore, $\NtD{j,0}$ and $\NtD{j,2}$ are $3 \times 3$ matrix operators satisfying
\begin{eqnarray}\label{NtD:j02}
  \NtD{j,0}{}
  \begin{bmatrix}
    \cmdpxu{j}{0}   \\
    \cmdpxu{j+1}{0} \\
    \cmdpyw{j}{1,-} \\
  \end{bmatrix}=
  \begin{bmatrix}
    \cmdu{j}{0}   \\
    \cmdu{j+1}{0} \\
    \cmdw{j}{1,-} \\
  \end{bmatrix},~
  \NtD{j,2}{}
  \begin{bmatrix}
    \cmdpyw{j}{2,+} \\
    \cmdpxu{j}{2}   \\
    \cmdpxu{j+1}{2} \\
  \end{bmatrix}=
  \begin{bmatrix}
    \cmdw{j}{2,+} \\
    \cmdu{j}{2}   \\
    \cmdu{j+1}{2} \\
  \end{bmatrix}.
\end{eqnarray}

To construct the NtD operators for $\Omega_{0}$ and $\Omega_{\NTC}$, there is no need to divide $\Omega_{0}$ and $\Omega_{\NTC}$ to subdomains. However, 
the field $u$ is discontinuous across the lines $x=x_{1}$ and $x=x_{\NTC}$ within the vertical interval $y_1<y<y_2$. This is because $u = u^{\ttot}$ for $x_1 < x < x_{\NTC}$, but $u = u^{\ttot} - \uref$ for $x < x_{1}$ and $x > x_{\NTC}$. In addition, $u=0$ on the three exterior edges of $\Omega_{0}$
and $\Omega_{\NTC}$. Therefore, $\NtD{0}$ and $\NtD{\NTC}$ are $3 \times 3$ matrix operators satisfying
\begin{eqnarray}
  \NtD{0}{}
  \begin{bmatrix}
    \cmdpxu{1}{0}                \\
    \cmdpxu{1}{1}-\cmdpxui{1}{1} \\
    \cmdpxu{1}{2}                \\
  \end{bmatrix}=\begin{bmatrix}
    \cmdu{1}{0}              \\
    \cmdu{1}{1}-\cmdui{1}{1} \\
    \cmdu{1}{2}              \\
  \end{bmatrix},~
  \NtD{\NTC}
  \begin{bmatrix}
    \cmdpxu{\NTC}{0}                   \\
    \cmdpxu{\NTC}{1}-\cmdpxui{\NTC}{1} \\
    \cmdpxu{\NTC}{2}                   \\
  \end{bmatrix}=\begin{bmatrix}
    \cmdu{\NTC}{0}                 \\
    \cmdu{\NTC}{1}-\cmdui{\NTC}{1} \\
    \cmdu{\NTC}{2}                 \\
  \end{bmatrix},\label{NtD:LRPML}
\end{eqnarray}
where $\cmdu{1}{1}=u(x_{1}^{+},y)$, $\cmdpxu{1}{1}=\cmdpxu(x_{1}^{+},y)$, $\cmdui{j}{1}=\uref(x_{1},y)$, $\cmdpxui{j}{1}=\partial_{x}\uref(x_{j},y)$ for $y_1<y<y_2$. Similarly, $\cmdu{\NTC}{1}$ and other
quantities with subscript $\NTC$ are obtained at $x=x_{\NTC}^{-}$. From Eqs.
\eqref{NtD:LRPML}, we easily obtain
\begin{eqnarray}
  \RTM{0}{}=
  \begin{bmatrix}
    0            \\
    \cmdui{1}{1} \\
    0            \\
  \end{bmatrix}- \NtD{0}{}\begin{bmatrix}
    0              \\
    \cmdpxui{1}{1} \\
    0              \\
  \end{bmatrix},~~\RTM{\NTC}{}=
  \begin{bmatrix}
    0               \\
    \cmdui{\NTC}{1} \\
    0               \\
  \end{bmatrix}- \NtD{\NTC}
  \begin{bmatrix}
    0                 \\
    \cmdpxui{\NTC}{1} \\
    0                 \\
  \end{bmatrix}.\label{RHS:0andT}
\end{eqnarray}

As defined in Eq. \eqref{Definition:u}, we know
that $u$ is discontinuous at the interface between the PML region and the
rectangle $(x_1,x_{\NTC})\times(y_1,y_2)$, where its jump is given by $\uref$.
Therefore,
\begin{eqnarray}
  \label{waveu:wj}
  \begin{aligned}
    \cmdw{j}{1,-}=\cmdw{j}{1,+}-\cmdwi{j}{1}, ~
    \cmdpyw{j}{1,-}=\cmdpyw{j}{1,+}-\cmdpywi{j}{1}, \\
    \cmdw{j}{2,+}=\cmdw{j}{2,-}-\cmdwi{j}{2}, ~\cmdpyw{j}{2,+}=\cmdpyw{j}{2,-}-\cmdpywi{j}{2},
  \end{aligned}
\end{eqnarray}
where $j=1,2,\cdots,\NTC-1$.

Using the above equations, we can construct $\NtD{j}$ and $\RTM{j}{}$. This
process involves two steps. First, we evaluate $\cmdw{j}{2,-}$ and
$\cmdw{j}{1,+}$ from Eqs. \eqref{NtD:j1} and \eqref{NtD:j02}, and substitute
into Eqs. \eqref{waveu:wj}. Second, we solve $\cmdpyw{j}{2,-}$ and $\cmdpyw{j}{1,+}$, substitute into Eqs. \eqref{NtD:j1} and \eqref{NtD:j02}. This leads to a linear relation between $\cmdu{j}$, $\cmdu{j+1}$, their $x$-derivatives, that is Eq. \eqref{NtD:Omegaj}. The inhomogeneous term
$\RTM{j}$ is related to $\uref$. Complete details are provided in Appendix
\ref{APP:Omegaj}.

\section{Matrix approximation for NtD operators}\label{Sec:approximation}

Since the entries of the coefficient matrix of Eq. \eqref{Eqn:FinalTri} are operators, and the elements of the unknown and right-hand-side vectors are functions of $y$, they must be approximated by matrices and vectors, respectively. There are two approaches to accomplish this. The classical approach
is to discretize $y$ by $K$ points, approximate each function of
$y$ by a column vector of length $K$, and approximate each operator by a
$K\times K$ matrix. Typically, this approach gives rise to a relatively large $K$, so that it is relatively expensive to solve Eq. \eqref{Eqn:FinalTri}. An
alternative approach is to approximate each function of $y$ by a sum of
properly chosen basis functions, and replace Eq. \eqref{Eqn:FinalTri} by a new
equation for the expansion coefficients, where the operators in Eq.
\eqref{Eqn:FinalTri} become matrices relating the coefficients. By taking
advantage of the piecewise smoothness of the wave field, the new equation can
be much smaller, leading to a more efficient numerical method.

We adopt the second approach, and approximate each function of $y$ the intervals $(y_0,y_1)$, $(y_1,y_2)$ and $(y_2,y_3)$ separately. Specifically, for the interval $(y_2,y_3)$, we use the following basis functions:
\begin{eqnarray*}
  \varphi_{2,m}(y)= (1-t)\exp(-C_{2}^{\upvarphi} t^2) L_{m-1}(2t-1), ~~ t=(y-y_2)/(y_3-y_2), ~~m=1,~2,~\cdots,~N_2,
\end{eqnarray*}
where $L_{m-1}$ is the Legendre polynomial of degree $m-1$, $C_{2}^{\upvarphi}$ is a positive constant, and $N_2$ is the total number of basis functions. For the interval $(y_0,y_1)$, the basis functions are
\begin{eqnarray*}
  \varphi_{0,m}(y)= (1-t)\exp(-C_{0}^{\upvarphi} t^2) L_{m-1}(2t-1), ~~ t=(y-y_1)/(y_0-y_1), ~~m=1,~2,~\cdots,~N_0.
\end{eqnarray*}
For the interval $(y_1,y_2)$, we use scaled Legendre polynomials that map $(y_1,y_2)$ to
$(-1,1)$, denote them as $\varphi_{1,m}(y)$, for $m=1,2,\cdots,N_1$, where each polynomial has degree $m-1$. The total number of basis functions is
$N=N_0+N_1+N_2$. Therefore, each function of $y$ is associated with $N$ expansion coefficients.

To approximate the operators in Eq. \eqref{Eqn:FinalTri}, we require the
matrix approximations $\NtDM{j,k}$ for the NtD operator $\NtD{j,k}$, where
$k=0,1,2$. 
The operator $\NtD{j,k}$ maps the normal derivative of $u$ to $u$ itself on the boundary of $\Omega_{j}^{(k)}$; correspondingly, the matrix $\NtDM{j,k}$ maps their respective vectors of expansion coefficients.
In what follows, we adopt the convention that a tilde over a function denotes the column vector of its expansion coefficients, and a tilde over an operator denotes the matrix that acts on these coefficient vectors. The specific definitions of these vectors are provided in Appendix \ref{Approx:vectors}.

On the interval $(x_{j},x_{j+1})$, we approximate each function of $x$ by a sum
of $M_{j}$ scaled Legendre polynomials $\phi_{j,m}(x)$, $m=1,2,\cdots,M_{j}$, where each polynomial $\phi_{j,m}(x)$ maps $(x_{j},x_{j+1})$ to $(-1,1)$. Consequently, matrix
$\NtDM{j,0}$ is of dimension $(2N_0+M_j)\times(2N_0+M_j)$, $\NtDM{j,2}$ is 
$(2N_2+M_j)\times(2N_2+M_j)$, $\NtDM{j,1}$ is $(2M_j+2N_1)\times
  (2M_j+2N_1)$, and these matrices satisfy
\begin{eqnarray}\label{NTDM:j012}
  \NtDM{j,0}{}
  \begin{bmatrix}
    \cmdpxuh{j}{0}   \\
    \cmdpxuh{j+1}{0} \\
    \cmdpywh{j}{1,-} \\
  \end{bmatrix}=
  \begin{bmatrix}
    \cmduh{j}{0}   \\
    \cmduh{j+1}{0} \\
    \cmdwh{j}{1,-} \\
  \end{bmatrix},~
  \NtDM{j,1}{}
  \begin{bmatrix}
    \cmdpywh{j}{1,+} \\
    \cmdpxuh{j}{1}   \\
    \cmdpxuh{j+1}{1} \\
    \cmdpywh{j}{2,-} \\
  \end{bmatrix}=\begin{bmatrix}
    \cmdwh{j}{1,+} \\
    \cmduh{j}{1}   \\
    \cmduh{j+1}{1} \\
    \cmdwh{j}{2,-} \\
  \end{bmatrix},~
  \NtDM{j,2}{}
  \begin{bmatrix}
    \cmdpywh{j}{2,+} \\
    \cmdpxuh{j}{2}   \\
    \cmdpxuh{j+1}{2} \\
  \end{bmatrix}=
  \begin{bmatrix}
    \cmdwh{j}{2,+} \\
    \cmduh{j}{2}   \\
    \cmduh{j+1}{2} \\
  \end{bmatrix}.
\end{eqnarray}
In addition, the matrices $\NtDM{j,k}$ ($k=0,1,2$) can be partitioned into blocks, they are given by $\NtDM{j,0}=[\NtDM{j,0}{m,l}]_{3\times 3}$, $\NtDM{j,1}=[\NtDM{j,1}{m,l}]_{4\times 4}$, $\NtDM{j,2}=[\NtDM{j,2}{m,l}]_{3\times 3}$, where $\NtDM{j,k}{m,l}$ is the $(m,l)$th block of $\NtDM{j,k}$.

For the NtD operators $\NtD{0}$ and $\NtD{\NTC}$, the approximations for
$\cmdu{j}{k}$ and $\cmdpxu{j}{k}$ ($j=1,\NTC$, $0\leq k\leq 2$) follow the
procedure outlined above. Additionally, we need the functions $\cmdui{j}{1}$ and $\cmdpxui{j}{1}$ ($j=1, \NTC$) on the interval $(y_1,y_2)$, which are also approximated by sums of $N_1$ scaled Legendre polynomials $\varphi_{1,m}(y)$. Consequently, $\NtDM{0}$, $\NtDM{\NTC}$
are $N\times N$ matrices that satisfy
\begin{eqnarray}
  \NtDM{0}{}
  \begin{bmatrix}
    \cmdpxuh{1}{0}                 \\
    \cmdpxuh{1}{1}-\cmdpxuih{1}{1} \\
    \cmdpxuh{1}{2}                 \\
  \end{bmatrix}=\begin{bmatrix}
    \cmduh{1}{0}               \\
    \cmduh{1}{1}-\cmduih{1}{1} \\
    \cmduh{1}{2}               \\
  \end{bmatrix},~
  \NtDM{\NTC}{}
  \begin{bmatrix}
    \cmdpxuh{\NTC}{0}                    \\
    \cmdpxuh{\NTC}{1}-\cmdpxuih{\NTC}{1} \\
    \cmdpxuh{\NTC}{2}                    \\
  \end{bmatrix}=\begin{bmatrix}
    \cmduh{\NTC}{0}                  \\
    \cmduh{\NTC}{1}-\cmduih{\NTC}{1} \\
    \cmduh{\NTC}{2}                  \\
  \end{bmatrix}.\label{NtDM:0andT}
\end{eqnarray}
Moreover, $\NtDM{0}$ and $\NtDM{\NTC}$ admit a $3\times 3$ block structure, with their $(m,l)$-th block denoted by $\NtDM{j}{m,l}$ (for $j=0,\NTC$ and $1\leq m,l\leq 3$).

We first discuss the steps for calculating the NtD matrix $\NtDM{j,1}$. 
According to the approximations for functions $\cmdpyw{j}{1,+}$,
$\cmdpyw{j}{2,-}$, $\cmdpxu{j}{1}$ and $\cmdpxu{j+1}{1}$ (see Eqs.
\eqref{approx:wj} and \eqref{approx:uij} in Appendix \ref{Approx:vectors}), we
need to solve $2(M_{j}+N_1)$ boundary value problems (BVPs) in subdomain
$\domain{j}{1}$. These BVPs are given by the Helmholtz equation
\eqref{Helmholtz:eqn} with different Neumann boundary conditions (NBCs). For
each BVP, the NBC is inhomogeneous on one edge and is zero on all other edges.
More precisely, the inhomogeneous NBC specifies that the $x$ or $y$ derivative
of the solution equals $\varphi_{1,m}(y)$ or $\phi_{j,m}(x)$, respectively.
Those BVPs are ordered by edge ``bottom, left, right, top'' and then by basis function index ($m=1,2,\cdots,N_1$ or $M_j$). We use FEM to
solve these BVPs. The corresponding linear systems have the same coefficient
matrix and different right-hand side vectors. Therefore, these linear systems
can be solved together efficiently. Let $\ulist$ be the solution of a
particular BVP. We expand the trace of $\ulist$ on the four edges as
\begin{eqnarray*}
  \ulist(x,y_{1})\approx\lsum{m=1}{M_{j}}\ulist{1}{m}\phi_{j,m}(x),&
  \ulist(x,y_{2})\approx\lsum{m=1}{M_{j}}\ulist{4}{m}\phi_{j,m}(x),&x\in(x_{j},x_{j+1}),\\
  \ulist(x_{j},y)\approx\lsum{m=1}{N_1}\ulist{2}{m}\varphi_{1,m}(y),&
  \ulist(x_{j+1},y)\approx\lsum{m=1}{N_1}\ulist{3}{m}\varphi_{1,m}(y),&y\in(y_{1},y_{2}).
\end{eqnarray*}
The coefficients are arranged into a column vector as follows:
\begin{eqnarray*}
  \left[
    \ulist{1}{1},\cdots,\ulist{1}{M_{j}},
    \ulist{2}{1},\cdots,\ulist{2}{N_1},
    \ulist{3}{1},\cdots,\ulist{3}{N_1},
    \ulist{4}{1},\cdots,\ulist{4}{M_{j}}
    \right]^{\top}.
\end{eqnarray*}
This vector itself forms a column of the matrix $\NtDM{j,1}$, ordered according to the sequence prescribed to the BVPs.

Next, we describe the steps for calculating the matrix $\NtDM{j,0}$. The two
functions $\cmdpxu{j}{0}$ and $\cmdpxu{j+1}{0}$ are approximated by sums of
$N_0$ basis functions $\varphi_{0,m}(y)$, and $\cmdpyw{j}{1,-}$ is approximated
by a sum of $M_{j}$ scaled Legendre polynomials $\phi_{j,m}(x)$. Similar to the
matrix $\NtDM{j,1}$, we need to solve $(2N_0+M_{j})$ BVPs in
$\domain{j}{0}$. These BVPs are given by Eq. \eqref{Helmholtz:PML} with
homogeneous Dirichlet boundary condition imposed on the bottom edge and
different NBCs imposed on all other edges. For each BVP, the NBC is nonzero on
one edge and is zero on other two edges. More precisely, the nonzero NBC
specifies that the $x$ or $y$ derivative of the solution equals
$\varphi_{0,m}(y)$ or $\phi_{j,m}(x)$, respectively. All these BVPs are ordered
first by edge ``left, right, top", then for each edge by the basis
functions $m=1,2,\cdots,N_0$ or $M_j$. Same as before, we expand the solutions
of these BVPs at the three edges (left, right, top), and the corresponding
expansion coefficients give rise to the entries of the matrix $\NtDM{j,0}$. The
matrix $\NtDM{j,2}$ can be calculated by the same method, with the
corresponding BVPs ordered by edge ``bottom, left, right''.

Finally, we describe the steps for calculating the matrix $\NtDM{0}$. The three
functions $\cmdpxu{1}{k}$ ($0\leq k\leq 2$) are approximated by sums of $N_{k}$
functions $\varphi_{k,m}(y)$, respectively, and $\cmdpxui{1}{1}$ is
approximated by a sum of $N_1$ scaled Legendre polynomials $\varphi_{1,m}(y)$.
Therefore, similar to the procedure outlined above, we need to solve $N$ BVPs
in $\Omega_{0}$. These BVPs are given by Eq. \eqref{Helmholtz:PML} with
different NBCs imposed on the right edge and homogeneous Dirichlet boundary
condition imposed on all other edges. These NBCs are also piecewisely defined.
For each BVP, the $x$ derivative of the solution on sub-interval
$(y_k,y_{k+1})$ of the right edge equals $\varphi_{k,1}(y)$ and is zero on the remaining parts.
For each BVP, the solution on the interval 
$(y_{k},y_{k+1})$ can be expanded in the same way as $\cmdpxu{1}{k}$, where
$k=0,1,2$. The expansion coefficients form a column of matrix $\NtDM{0}$. The
NtD matrix $\NtDM{\NTC}$ can be calculated similarly.

Using the NtD matrices for the subdomains $\Omega_{j}^{(k)}$, $k=0,1,2$, and
following the procedure given in Sec. 3, we compute the matrix $\NtDM{j}$
and the source vector $\RTMh{j}$ for $\RTM{j}$.
For this step, the continuity conditions at $y=y_1$ and $y=y_2$ are given as
\begin{eqnarray}\label{DisCon:discrete}
  \begin{aligned}
    \cmdwh{j}{1,-}=\cmdwh{j}{1,+}-\cmdwih{j}{1},~~~
    \cmdpywh{j}{1,-}=\cmdpywh{j}{1,+}-\cmdpywih{j}{1}, \\
    \cmdwh{j}{2,+}=\cmdwh{j}{2,-}-\cmdwih{j}{2},~~~
    \cmdpywh{j}{2,+}=\cmdpywh{j}{2,-}-\cmdpywih{j}{2},
  \end{aligned}
\end{eqnarray}
where $\cmdwih{j}{k}$ ($k=1,\ 2$) are the expansion coefficients of $\cmdwi{j}{k}$.

By following the discretization procedure for Eq. \eqref{Eqn:FinalTri}, we obtain a block
tridiagonal matrix linear system that we denote as (9A), which has the same
form as Eq. \eqref{Eqn:FinalTri} with
every quantity (entries of the
coefficient matrix and the elements of the unknown and right-hand-side vectors)
replaced by its tilde-adorned discrete counterpart. 
Equation (9A) can be solved using block LU factorization
\cite{2020-Lyche-Numerical}, the complexity of which is $O(N^3\NTC)$. Its solution yields $\cmdpywh{j}{1,+}$ and
$\cmdpywh{j}{2,-}$ following the procedure given in Appendix \ref{APP:Omegaj}. Then, we use Eq. \eqref{DisCon:discrete} to obtain $\cmdpywh{j}{1,-}$ and
$\cmdpywh{j}{2,+}$. 
Finally, the wave field in each subdomain is reconstructed as a linear combination of the BVP solutions used to build the local NtD matrix, with the combination weights directly provided by the solution of system (9A).

\section{Far-field by plane wave expansion}\label{sec:far-field}
Using the method described in the previous section, we obtain the wave field
within a bounded domain $(x_{1},x_{\NTC})\times(y_1,y_2)$. However, the far-field is often of primary interest. To recover the far-field for $y>y_2$ and $y<y_1$, we
first recover the data on the lines $y=y_1$ and $y=y_2$ for $x<x_1$ and
$x>x_{\NTC}$ using data from the PML region,
then compute the Fourier transform of this data along lines $y=y_1$ and $y_2$, and
finally apply the inverse Fourier transform to recover the far-field.

For $y> y_2$, the scattering field $u^{(\text{s})}=u^{\ttot}-\uref$ satisfies
Eq. \eqref{Helmholtz:eqn} for refractive index $n = 1$. Applying the Fourier
transform to Eq. \eqref{Helmholtz:eqn} with respect to $x$, we obtain
\begin{eqnarray}
  \label{Helmholtz:Eqn_frequencyx}
  \frac{\partial^2 \widehat{u}^{(\text{s})}(\alpha,y)}{\partial y^2}+\beta^2 \widehat{u}^{(\text{s})}(\alpha,y)=0,
  ~~\beta=\beta(\alpha)=\sqrt{\wavel^2-\alpha^2}.
\end{eqnarray}
The Fourier transform of $u^{(\text{s})}(x,y_2)$ is
\begin{eqnarray}
  \begin{aligned}
    \widehat{u}^{(\text{s})}(\alpha,y_2)=  \int_{-\infty}^{\infty}u^{(\text{s})}(x,y_2)\exp(-i\alpha x)dx
    =                               \left(\int_{-\infty}^{x_1}+\int_{x_1}^{x_{\NTC}}+\int_{x_{\NTC}}^{\infty}\right)u^{(\text{s})}(x,y_2)\exp(-i\alpha x)dx.
  \end{aligned}
\end{eqnarray}
Since $u^{(\text{s})}$ is directly available at $y=y_2$ for $x< x_{1}$ and $x> x_{\NTC}$, the first and third integrals above should be transformed into integrals along the complex paths defined by the PML. Accordingly, they become:
\begin{eqnarray*}
  \begin{aligned}
    \int_{-\infty}^{x_1}u^{(\text{s})}(x,y_2)\exp(-i\alpha x)dx=
    \int_{x_{0}}^{x_{1}}u^{(\text{s})}_{\text{PML}}(x,y_2)\exp(-i\alpha \hat{x})s_x(x)dx,
  \end{aligned}\\
  \begin{aligned}
    \int_{x_{\NTC}}^{\infty}u^{(\text{s})}(x,y_2)\exp(-i\alpha x)dx=
    \int_{x_{\NTC}}^{x_{\NTC+1}}u^{(\text{s})}_{\text{PML}}(x,y_2)\exp(-i\alpha \hat{x})s_x(x)dx,
  \end{aligned}
\end{eqnarray*}
where $u^{(\text{s})}_{\text{PML}}(x,y_2)$ is the PML-transformed revision of $u^{\ttot}-\uref$. It is denoted as $u$ in Sec. 2, and satisfies Eq. \eqref{Helmholtz:PML}.

Since the scattering field $u^{(\text{s})}$ can only propagate upwards for $y > y_2$, we have
\begin{eqnarray*}
  \widehat{u}^{(\text{s})}(\alpha,y)=\widehat{u}^{(\text{s})}(\alpha,y_2)\exp(i\beta (y-y_2)).
\end{eqnarray*}
Using the inverse Fourier transform, we obtain the far field for $y>y_2$ as follows
\begin{eqnarray}\label{Far-field:uxy}
  u^{(\text{s})}(x,y)=\frac{1}{2\pi}\int_{-\infty}^{\infty}\widehat{u}^{(\text{s})}(\alpha,y_2)\exp(i\beta(\alpha) (y-y_2))\exp(i\alpha x)d\alpha.
\end{eqnarray}

From the expression $\beta=\sqrt{\wavel^2-\alpha^2}$, we note that $\beta$ is a
pure imaginary number when $\alpha>\wavel$;  consequently, Eq. \eqref{Far-field:uxy}
can be rewritten as
\begin{eqnarray}\label{u_in_far_field}
  \begin{aligned}
    u^{(\text{s})}(x,y)= & \frac{1}{2\pi}\int_{-\alpha}^{\alpha}\widehat{u}^{(\text{s})}(\alpha,y_2)\exp(i\beta(\alpha) (y-y_2))\exp(i\alpha x)d\alpha                                        \\
    +             & \frac{1}{2\pi}\left(\int_{-\infty}^{-\alpha}+\int_{\alpha}^{\infty}\right)\widehat{u}^{(\text{s})}(\alpha,y_2)\exp(-\gamma(\alpha) (y-y_2))\exp(i\alpha x)d\alpha,
  \end{aligned}
\end{eqnarray}
where $\gamma(\alpha)=\sqrt{\alpha^2-\wavel^2}$ is a non-negative real number.
It is obvious that the first and second parts of Eq. \eqref{u_in_far_field} are propagating and evanescent waves, respectively. The evanescent wave decays exponentially with the distance $y-y_2$ and therefore does not contribute to the far field.

\section{Numerical examples}
\begin{figure}[H]
  \centering
  \begin{tikzpicture}[scale=2.2]
    \def\height{0.5}
    \def\WL{0.2}
    \def\xM{7}
    \def\yp{{-0.2,0,0.9,1.35,2.7,2.9}}
    \filldraw[green] (0,\yp[2]) rectangle (\xM,\yp[3]);
    \draw[black,-latex,thick] (3,\yp[4]-0.3) -- (3,\yp[4]-0.8);
    \node at (3.5,\yp[4]-0.5) {\small Incoming wave};
    \foreach \i in {2,3}{
        \draw[blue,thin] (0,\yp[\i]) -- (\xM,\yp[\i]);
      }
    \node at (\xM/2,\yp[2]*0.5+\yp[3]*0.5) {Slab};
    \foreach \k in {0,1,2}{
        \foreach \i in {1,...,7}{
            \pgfmathparse{sqrt(\i+6)};
            \def\wj{\WL*\pgfmathresult/8};
            \def\xj{1.6*\k+\i*\WL};
            \coordinate (P1) at (\xj+\wj,\yp[3]);
            \coordinate (P3) at (\xj+\WL-\wj,\yp[3]+0.5);
            \filldraw[red] (P1) rectangle (P3);
            \draw (P1) rectangle (P3);
          }
      }
    \node at (5.3,\yp[3]+0.3) {$\cdots$};
    \foreach \i in {3,...,6,7}{
        \pgfmathparse{sqrt(\i+6)};
        \def\wj{\WL*\pgfmathresult/8};
        \def\xj{5.2+\i*\WL};
        \coordinate (P1) at (\xj+\wj,\yp[3]);
        \coordinate (P3) at (\xj+\WL-\wj,\yp[3]+0.5);
        \filldraw[red] (P1) rectangle (P3);
        \draw (P1) rectangle (P3);
      }
  \end{tikzpicture}
  \caption{A sketch of 2D phase gradient metasurface.}\label{MS:phase_gradient}
\end{figure}
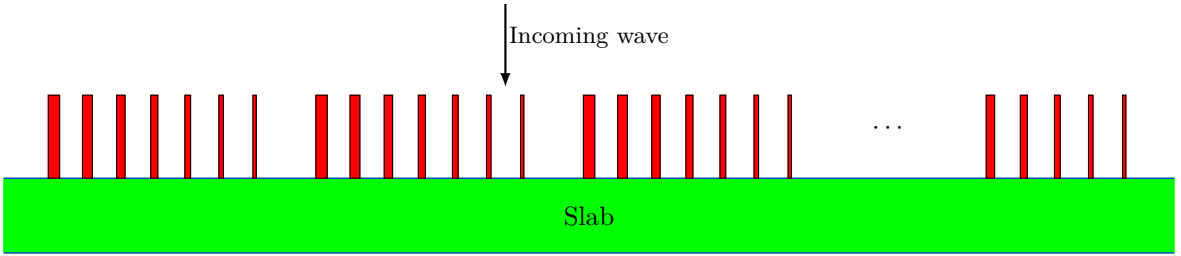

In this section, we use a phase gradient metasurface \cite{2014-Lin-S} shown in Fig.
\ref{MS:phase_gradient} as an example to verify the accuracy and demonstrate
the efficiency of our method. It is a 2D metasurface constructed following the
approach used in Ref. \cite{2020-Lawrence-NN}, and it
consists of some rectangular cylinders placed on an infinite synthetic sapphire
($\alpha\text{-Al}_2\text{O}_3$) slab with refractive index
$n_{\text{s}}=1.7743$ and a thickness of $0.45~\micron$. These rectangular
cylinders are made of Gallium nitride (GaN) with a refractive index
$n_{\text{c}}=2.4431$, a fixed height of $0.5~\micron$ in the $y$ direction and
varying widths in the $x$ direction. The refractive indices $n_{\text{s}}$ and
$n_{\text{c}}$ correspond to the free space wavelength $\lambda=0.5~\micron$
\cite{2024-Polyanskiy-Refractiveindex.info}.

The design utilizes $7$ distinct cylinder widths given by $0.2(1-\sqrt{j+6}/4)~\micron$,
$j=1,2,\cdots,7$. The metasurface consists of unit cells with a constant size
$L=0.2~\micron$ for the $x$ direction.
We define $x_j=(j-1)L$ for
$j=1,\cdots,\NTC$, then the $j$-th unit cell occupies the interval $(x_j,x_{j+1})$. More specifically, we consider a metasurface with a period of $8$
unit cells as shown in Fig. \ref{MS:phase_gradient}, where the first unit cell
$(x_1,x_2)$ is an empty one (no cylinder), and the $j$-th unit cell
$(x_j,x_{j+1})$ contains the cylinder with $w_{j-1}$ for $j=2,3,\cdots,8$.
Therefore, including the empty cell, there are $\NC=8$ distinct unit cells in
this device. The scattering problem is for a plane wave with wavelength
$\lambda=2\pi/k_0=0.5~\micron$ and a zero incident angle illuminating this
metasurface from above.

To validate our method, we first consider a small metasurface with only $7$ cylinders. The computational domain contains $9$ unit cells where the
first and last ones, i.e. $(x_1,x_2)$ and $(x_9,x_{10})$, are empty cells. The
thicknesses of the left and right PML regions are $0.2~\micron$. This implies that
$\NTC=10$, $x_0=-0.2~\micron$ and $x_{11}=2~\micron$. In the $y$ direction, we
truncate this structure as shown in Fig. \ref{Sketch:Metasurface}, and choose
$(y_0,y_1,y_{\text{sb}},y_{\text{st}},y_{2},y_{3})=(-0.2,0,0.6,1.05,2.1,2.3)~\micron$. For the PML regions, the positive constant $b$ is around $5.35\cdot 10^{5}$. We use our method and the classical FEM to simulate this device, and then
compare the results and calculate the relative error of the scattering field
$u^{(\text{s})}=u^{\text{tot}}-u^{\text{ref}}$ on the line $y=y_2$ for
$x_{1}<x<x_{\NTC}$. The relative error $\Delta$ is defined as
$\|u^{(\text{s})}_{\text{NtD}}(x,y_2)-u^{(\text{s})}_{\text{FEM}}(x,y_2)\|/\|u^{(\text{s})}_{\text{FEM}}(x,y_2)\|$,
where the norm is $\|u\|^2=\int_{x_1}^{x_{\NTC}}|u|^2dx$. For our method, we
choose the number of basis functions on different edges as $N_1=35$,
$N_0=N_2=8$, and $M_j=15$ ($j=1,2,\cdots,9$). The two positive constants in basis
functions $\varphi_{k,m}(y)$ are chosen as $C_{k}^{\upvarphi}=5$ ($k=0,2$). As a
well-tested numerical method, FEM is used for solving the scattering problem in
the rectangular domain $(x_{0},x_{\NTC})\times(y_0,y_3)$ with the same PML
parameters. The same reference mesh size $h$ is used for both FEM and our method.
It guarantees that the area of every triangular element is bounded by $h^2/2$.

We perform simulations at multiple mesh sizes $h=L/10$, $L/20$, $L/40$ and $L/80$ for both methods. The results are summarized in Tab.
\ref{Tab:Error}. For $h=L/10$, our method gives rise to $1966$, $1801$, $160$
and $167$ discretization points in $\domain{0}$, $\domain{\NTC}$, $\domain{j}{0}$ and
$\domain{j}{2}$, respectively. For the subdomain $\domain{j}{1}$ ($1\leq j <\NTC$), the number of points ranges from $1681$ to $1723$, resulting in a total of $19961$ points. For the same mesh size $h=L/10$, FEM yields $22028$ discretization
points. The corresponding relative error is $\Delta_{L/10}\approx 0.054875$. The data indicates that the
relative error $\Delta$ on a mesh of size $h$ is approximately a quarter of
that on a coarser mesh of size $2h$. This shows that the solution obtained by our method converges to the FEM benchmark at a second-order rate, as quantified by $\log_2(\Delta_{2h}/\Delta_{h})$. Since FEM with linear elements is of second order
accuracy, this result confirms that our method also achieves second-order accuracy.

\begin{table}[H]
  \setlength{\abovecaptionskip}{2pt}
  \setlength{\belowcaptionskip}{3pt}
  \centering
  \caption{The information of meshes and relative errors}\label{Tab:Error}
\begin{tabular}{cccccccccc}
    \toprule
    
    $h$ & $\domain{0}$ & $\domain{\NTC}$ & $\domain{j}{0}$ & $\domain{j}{1}$  & $\domain{j}{2}$ & Total   & FEM     & Error $\Delta$ & Rate \\\midrule
    $L/10$        & 1966         & 1801            & 160              & $[1681, 1723]$    & 167              & 19961   & 22028   & 0.054875       & -    \\\midrule
    $L/20$        & 7602         & 7295            & 643              & $[6464,6528]$     & 643              & 77218   & 87506   & 0.014105       & 1.96 \\\midrule
    $L/40$        & 30177        & 29379           & 2461             & $[25386,25523]$   & 2456             & 302454  & 349131  & 0.003431       & 2.04 \\\midrule
    $L/80$        & 119735       & 118230          & 9586             & $[100500,100753]$ & 9592             & 1196214 & 1393938 & 0.000869       & 1.98 \\\bottomrule
  \end{tabular}
\end{table}

\begin{figure}[H]
  \centering
  \subfigure[$h=L/10$]{\includegraphics[width=7.8cm]{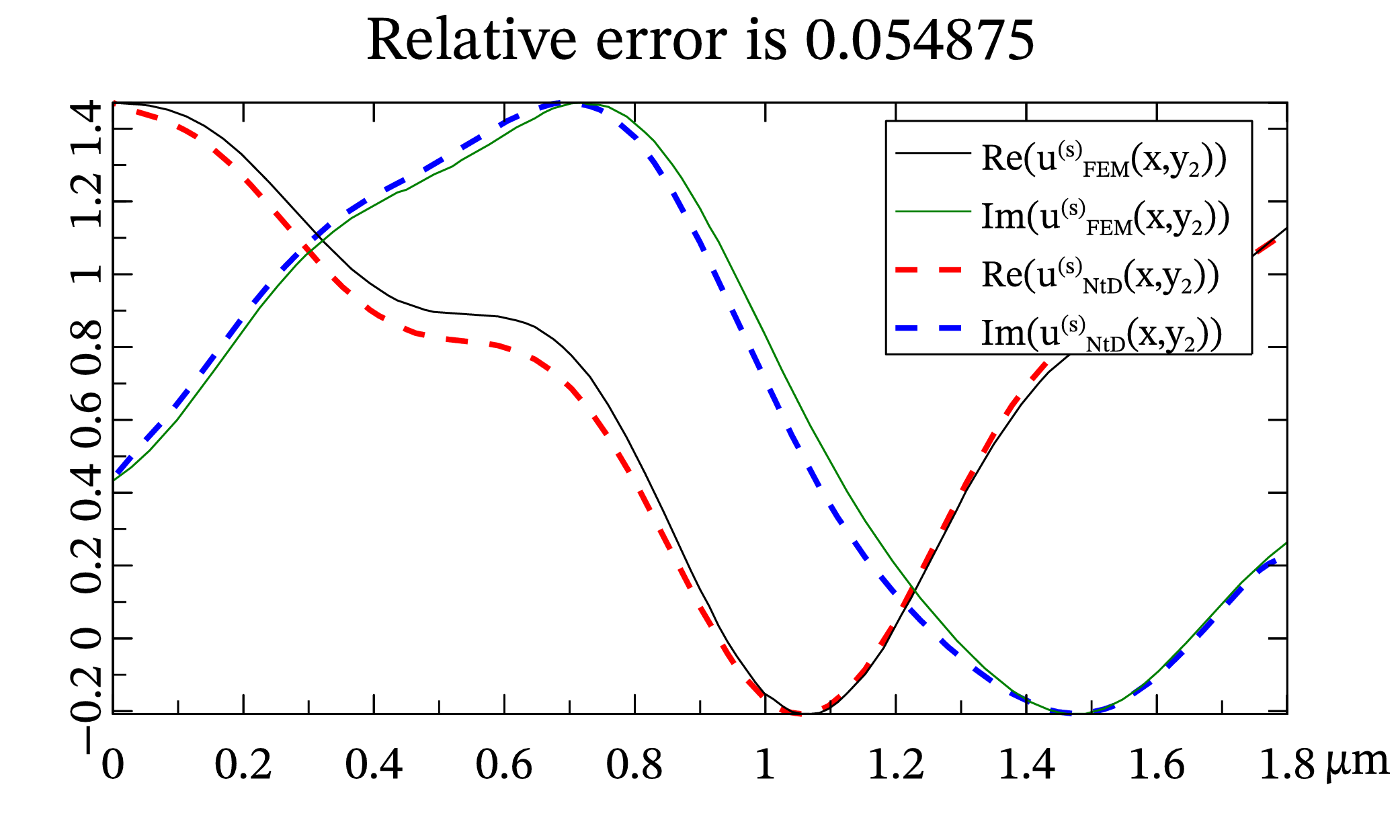}}
  \subfigure[$h=L/40$]{\includegraphics[width=7.8cm]{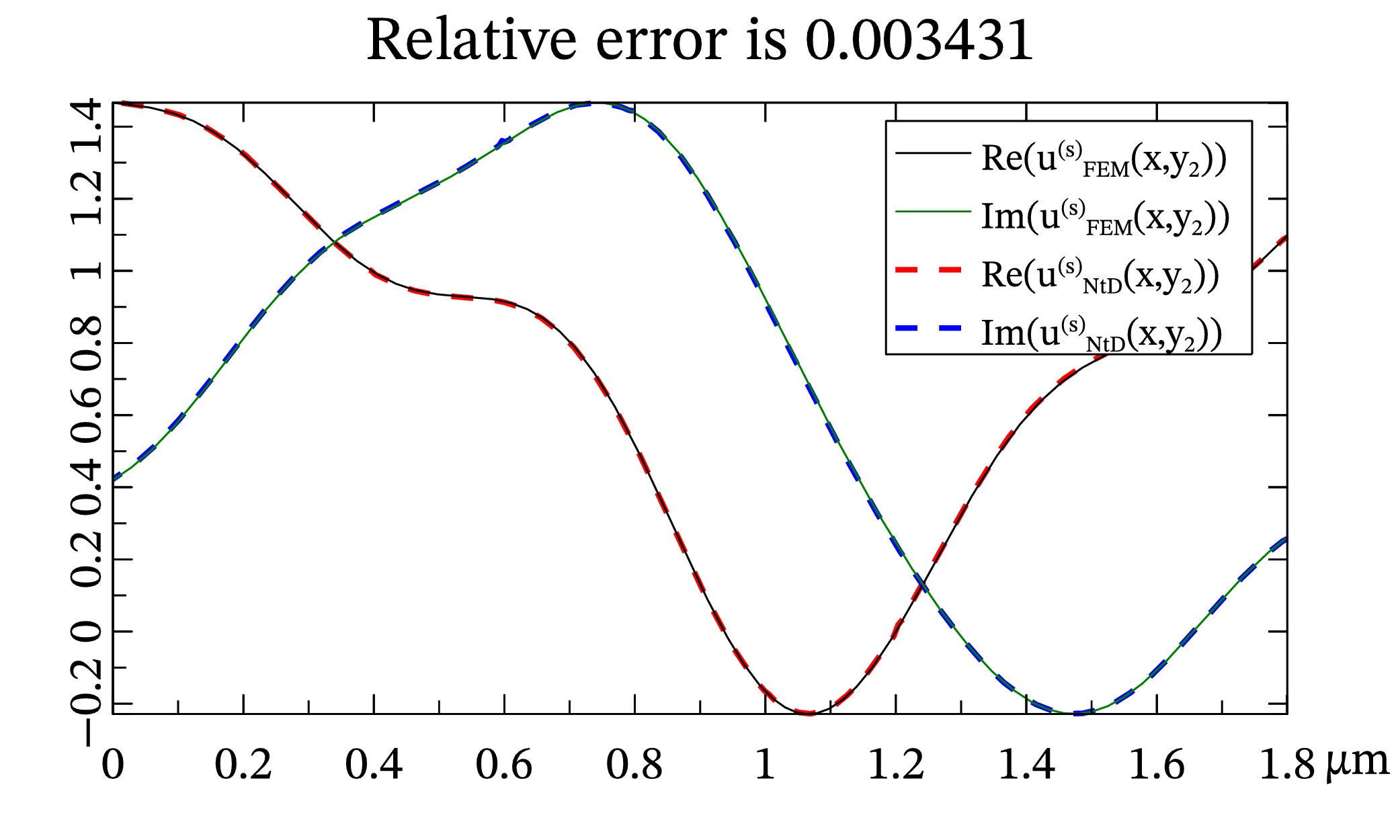}}
  \caption{Relative errors of the numerical solutions obtained by our method and FEM.}\label{Fig:uhy2}
\end{figure}

\begin{figure}
  \centering
  \subfigure[Our method]{\includegraphics[height=7.0cm]{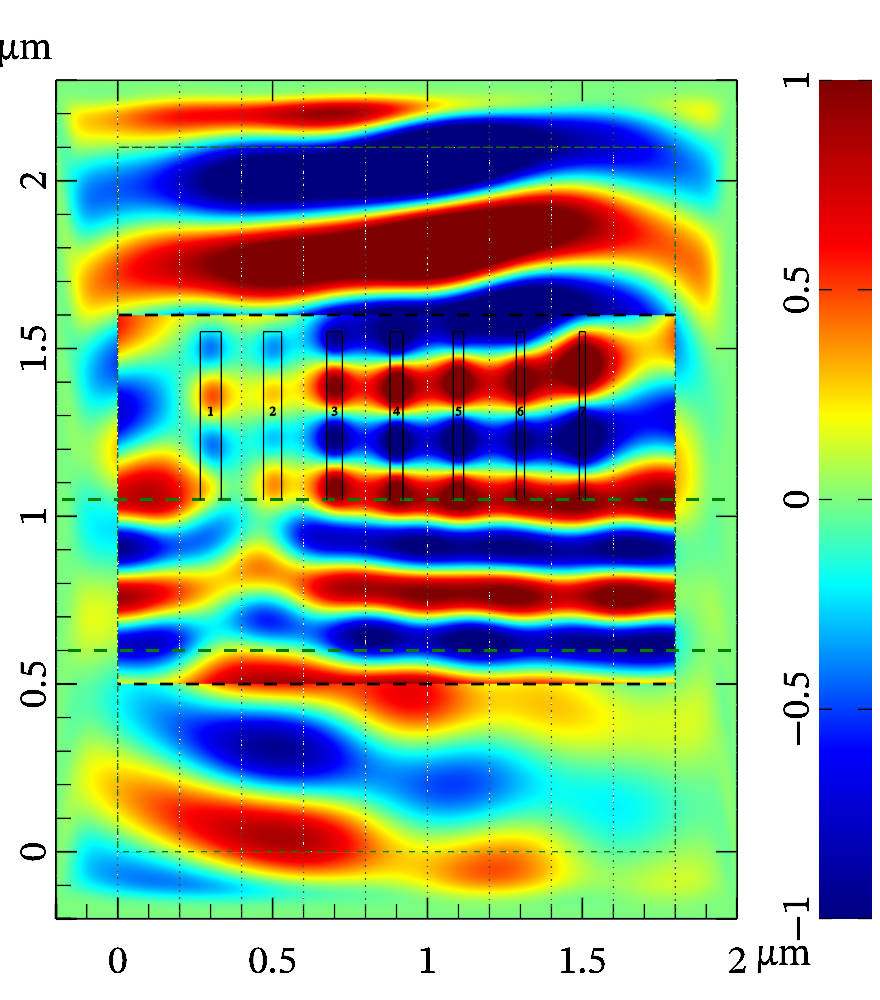}\label{Fig:NtDresult}}
  \subfigure[FEM]{\includegraphics[height=7.0cm]{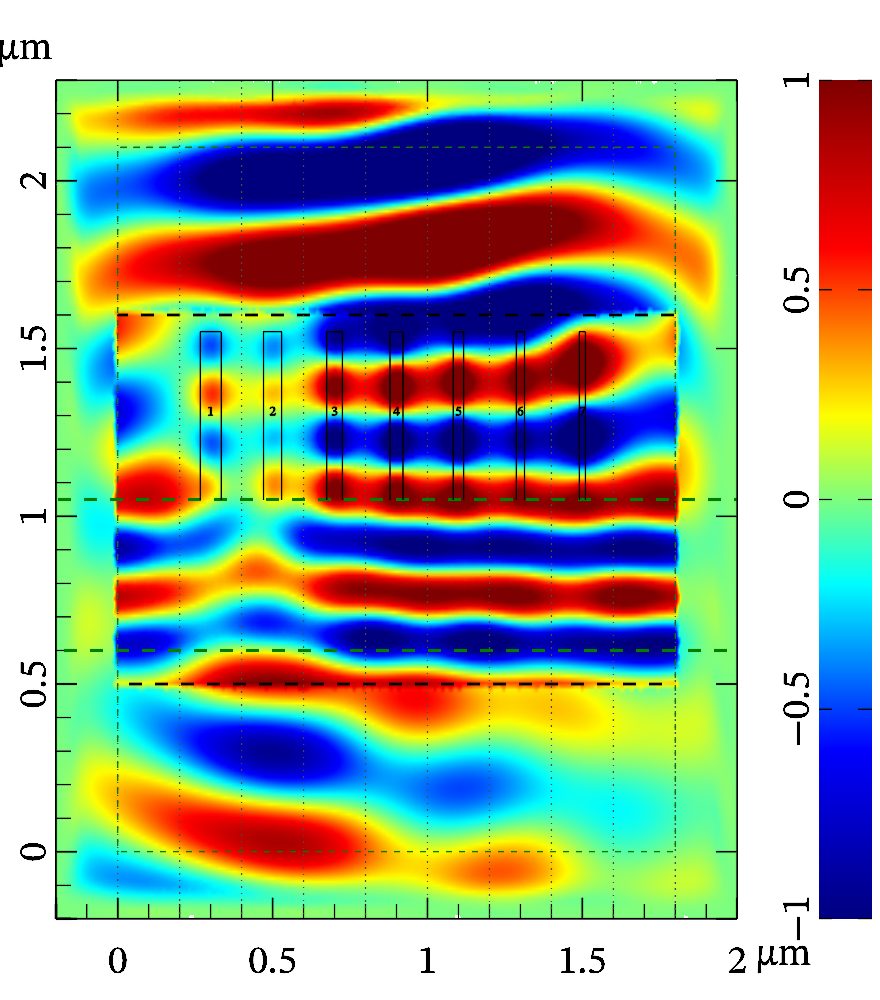}\label{Fig:FEMresult}}
  \caption{Simulations for the phase gradient metasurface solved by our method (Left) and FEM (Right).}\label{Fig:PW:ref_T}
\end{figure}

We compare the scattering fields computed by both methods along the line $y=y_2$ for $x_1 < x < x_{\NTC}$ in Fig. \ref{Fig:uhy2}. Figure \ref{Fig:PW:ref_T} presents the full numerical solution obtained with the reference mesh size $h = L/80$. It displays the total field $u^{\text{tot}}$ within the rectangular domain $(0, 1.8) \times (0.5, 1.6)~\mu\text{m}$ and the scattered field $u^{(\text{s})}$ in the surrounding region.

Next, to demonstrate the efficiency of our method, we simulate the metasurface
with a much larger number of cells by both FEM and our method. From the
previous section, it is clear that the NtD matrices can be calculated
independently. We develop two separate implementations for our method which
calculate the NtD matrices serially and in parallel, respectively. The
numerical results are obtained for reference mesh size $h=L/20$. To have a fair
comparison, we have calculated the solution on the entire computational domain
for both methods. Details of the mesh used in our method are given in Table \ref{Tab:Error}. For FEM, the number of mesh points is around
$7940(\NTC+1)$. 
Table \ref{Tab:Nx_20} summarizes the CPU time and peak memory usage for both methods, along with the corresponding relative error $\Delta$.
\begin{table}[H]
  \centering
  \setlength{\abovecaptionskip}{2pt}
  \setlength{\belowcaptionskip}{3pt}
  \caption{CPU time and peak memory required by these two methods.}\label{Tab:Nx_20}
  \begin{tabular}{r|ccc|ccc|c}
    \toprule
    \multirow{2}*{$\NTC-1$} & \multicolumn{3}{c|}{CPU time (seconds)} & \multicolumn{3}{c|}{Peak memory (GB)}                                                      &\multirow{2}*{$\Delta$}\\
    \cline{2-7}
                            & FEM                               & Serial NtD                           & Parallel NtD & FEM     & Serial NtD & Parallel NtD \\\midrule
    10                      & 8.52                              & 171.25                               & 28.86        & 0.7427  & 0.2430     & 0.5068    &0.013788   \\
    100                     & 82.2                              & 171.72                               & 29.42        & 5.5103  & 0.2433     & 0.4560       &0.013474\\
    200                     & 169.79                            & 175.01                               & 29.95        & 10.9329 & 0.2546     & 0.4751       &0.013730\\
    300                     & 265.39                            & 173.4                                & 30.59        & 15.9908 & 0.2643     & 0.4582       &0.013707\\
    400                     & 361.39                            & 173.31                               & 32.42        & 21.7391 & 0.2610     & 0.4629       &0.013637\\
    1000                    & /                                 & 176.21                               & 34.08        & /       & 0.3222     & 0.5317       &/\\
    10000                   & /                                 & 222.66                               & 79.85        & /       & 1.6891     & 1.8945       &/\\
    50000                   & /                                 & 438.68                               & 296.6        & /       & 7.7555     & 7.9761      &/ \\
    100000                  & /                                & 732.51                               & 593.75       & /       & 15.3538    & 15.5836     &/ \\
    150000                  & /                                 & 1087.90                              & 947.92       & /       & 22.9183    & 23.0845      &/\\
    \bottomrule
  \end{tabular}
\end{table}
As discussed in the previous section, our method first computes all NtD matrices, then calculates the normal derivatives of the wave field on the interfaces, and finally recovers the wave field within all subdomains. The computation of NtD matrices is only related to $\NC$, $N_{k}$ ($k=0,1,2$), $M_{j}$ ($1\leq j<\NTC$), and the mesh size $h$.
Table \ref{Tab:Nx_20} indicates that the NtD matrices can be computed in parallel, achieving a speedup of about $6$ on a personal computer with an 8-core CPU.
Moreover, the CPU time of the second and third steps and peak memory consumption grow at a much slower rate than those of a full domain FEM.
The parallel version of our method consumes significantly less CPU time (in the step for calculating the NtD matrices) compared to serial implementation, at the cost of only slightly more memory. The relative error does not change significantly with the increase in the number of unit cells.
From Tab. \ref{Tab:Nx_20}, it is clear that our method is highly efficient, advantageous and accurate for simulating large 2D metasurfaces.
All the numerical experiments are implemented in C++ and performed on a personal computer with an AMD CPU (8-core Ryzen 7 7840HS) and 32GB memory.

\section{Conclusion}

In this paper, we propose an NtD-based method for simulating 2D non-periodic metasurfaces, which consist of a huge number of subwavelength elements on a slab. The fundamental assumption is that there are only a relatively small number of distinct elements. This method relies on the NtD operators for subdomains of distinct unit cells and the PML regions. They are approximated by matrices and calculated using the FEM and local function expansions. In the first step, we assemble and solve a block tridiagonal linear system for the normal derivative of the wave field on the interfaces between the unit cells. The number of unknowns in this linear system is much less than that in the classical FEM. In the second step, we recover the wave field inside each unit cell and calculate the far-field, if necessary. Numerical examples indicate that our method is accurate and efficient for simulating large 2D metasurfaces with $10^5$ or more subwavelength elements.

\section*{Acknowledgements}
This work was supported by the Research Grants Council of Hong Kong Special Administrative Region, China (CityU 11317622).


\begin{thebibliography}{10}

\bibitem{2013-Kildishev-S}
A.~V. Kildishev, A.~Boltasseva, and V.~M. Shalaev, ``Planar photonics with metasurfaces,'' {\em Science}, vol.~339, no.~6125, 2013.

\bibitem{2017-Khorasaninejad-S}
M.~Khorasaninejad and F.~Capasso, ``Metalenses: Versatile multifunctional photonic components,'' {\em Science}, vol.~358, no.~6367, 2017.

\bibitem{2025-Brongersma-NREE}
M.~L. Brongersma, R.~A. Pala, H.~Altug, F.~Capasso, W.~T. Chen, A.~Majumdar, and H.~A. Atwater, ``The second optical metasurface revolution: moving from science to technology,'' {\em Nat. Rev. Electr. Eng.}, vol.~2, no.~2, pp.~125--143, 2025.

\bibitem{2014-Yu-NM}
N.~Yu and F.~Capasso, ``Flat optics with designer metasurfaces,'' {\em Nat. Mater.}, vol.~13, no.~2, pp.~139--150, 2014.

\bibitem{2015-Ni-S}
X.~Ni, Z.~J. Wong, M.~Mrejen, Y.~Wang, and X.~Zhang, ``An ultrathin invisibility skin cloak for visible light,'' {\em Science}, vol.~349, no.~6254, pp.~1310--1314, 2015.

\bibitem{2016-Khorasaninejad-S}
M.~Khorasaninejad, W.~T. Chen, R.~C. Devlin, J.~Oh, A.~Y. Zhu, and F.~Capasso, ``Metalenses at visible wavelengths: Diffraction-limited focusing and subwavelength resolution imaging,'' {\em Science}, vol.~352, no.~6290, pp.~1190--1194, 2016.

\bibitem{2022-Santiago-Cruz-S}
T.~Santiago-Cruz, S.~D. Gennaro, O.~Mitrofanov, S.~Addamane, J.~Reno, I.~Brener, and M.~V. Chekhova, ``Resonant metasurfaces for generating complex quantum states,'' {\em Science}, vol.~377, no.~6609, pp.~991--995, 2022.

\bibitem{2025-Soma-NC}
G.~Soma, K.~Komatsu, Y.~Nakano, and T.~Tanemura, ``Complete vectorial optical mode converter using multi-layer metasurface,'' {\em Nat. Commun.}, vol.~16, no.~1, 2025.

\bibitem{2025-Gu-N}
S.~Gu, C.~Mao, A.~Guell~Izard, S.~Sadana, D.~Terrel-Perez, M.~Mettry-Yassa, W.~Choi, W.~Zhou, H.~Yan, Z.~Zhou, T.~Massey, A.~Abelson, Y.~Zhou, S.~Huang, C.~Daraio, T.~U. Tumkur, J.~A. Fan, and X.~Xia, ``3d nanolithography with metalens arrays and spatially adaptive illumination,'' {\em Nature}, vol.~648, no.~8094, pp.~591--599, 2025.

\bibitem{2017-Ding-Wave}
Y.~S. Ding and Y.~He, ``Wave-vector and polarization dependent impedance model for a hexagonal periodic metasurface exemplified through finite-difference time-domain simulations,'' {\em Opt. Express}, vol.~25, no.~17, p.~20757, 2017.

\bibitem{2025-Huang-JCAM}
Y.~Huang, C.~Li, and J.~Li, ``Developing and analyzing a {FDTD} method for simulation of metasurfaces,'' {\em J. Comput. Appl. Math.}, vol.~460, p.~116364, 2025.

\bibitem{2023-Chen-IToAaP}
L.~Chen, M.~B. Ozakin, R.~Zhao, and H.~Bagci, ``A locally-implicit discontinuous {G}alerkin time-domain method to simulate metasurfaces using generalized sheet transition conditions,'' {\em IEEE Trans. Antennas Propag.}, vol.~71, no.~1, pp.~869--881, 2023.

\bibitem{2025-Li-JoCP}
C.~Li, Y.~Huang, and J.~Li, ``Developing and analyzing a {DG} method for modeling of metasurfaces,'' {\em J. Comput. Phys.}, vol.~534, p.~114011, 2025.

\bibitem{2016-Vahabzadeh-ITAP}
Y.~Vahabzadeh, K.~Achouri, and C.~Caloz, ``Simulation of metasurfaces in finite difference techniques,'' {\em IEEE Trans. Antennas Propag.}, vol.~64, no.~11, pp.~4753--4759, 2016.

\bibitem{2017-Sandeep-ITAP}
S.~Sandeep, J.-M. Jin, and C.~Caloz, ``Finite-element modeling of metasurfaces with generalized sheet transition conditions,'' {\em IEEE Trans. Antennas Propag.}, vol.~65, no.~5, pp.~2413--2420, 2017.

\bibitem{2023-Yang-CPC}
W.~Yang, T.~Wang, and J.~Mao, ``Adaptive edge finite element method and numerical design for metasurface cloak,'' {\em Comput. Phys. Commun.}, vol.~292, p.~108858, 2023.

\bibitem{2023-Amboli-OE}
J.~Amboli, B.~Gallas, G.~Demésy, and N.~Bonod, ``Design and analysis of chiral and achiral metasurfaces with the finite element method,'' {\em Opt. Express}, vol.~31, no.~26, p.~43147, 2023.

\bibitem{2026-Mao-JoCaAM}
J.~Mao, W.~Wang, and W.~Yang, ``An adaptive finite element method based on generalized sheet transition conditions and its application to modeling and simulation of functional metasurfaces,'' {\em J. Comput. Appl. Math.}, vol.~474, p.~116882, 2026.

\bibitem{2015-Francavilla-IToAaP}
M.~A. Francavilla, E.~Martini, S.~Maci, and G.~Vecchi, ``On the numerical simulation of metasurfaces with impedance boundary condition integral equations,'' {\em IEEE Trans. Antennas Propag.}, vol.~63, no.~5, pp.~2153--2161, 2015.

\bibitem{2023-Cai-IToAaP}
G.~Cai, X.~Liu, T.~Shen, J.~Liu, N.~Liu, and Q.~H. Liu, ``A full-vectorial spectral element method with generalized sheet transition conditions for high-efficiency metasurface/metafilm simulation,'' {\em IEEE Trans. Antennas Propag.}, vol.~71, no.~3, pp.~2652--2660, 2023.

\bibitem{2009-Dolean-SJoSC}
V.~Dolean, M.~J. Gander, and L.~Gerardo-Giorda, ``Optimized {S}chwarz methods for {M}axwell's equations,'' {\em SIAM J. Sci. Comput.}, vol.~31, no.~3, pp.~2193--2213, 2009.

\bibitem{2013-Chen-SJoNA}
Z.~Chen and X.~Xiang, ``A source transfer domain decomposition method for {H}elmholtz equations in unbounded domain,'' {\em SIAM J. Numer. Anal.}, vol.~51, no.~4, pp.~2331--2356, 2013.

\bibitem{2016-Tao-MC}
S.~Tao, J.~Cheng, and H.~Mosallaei, ``An integral equation based domain decomposition method for solving large-size substrate-supported aperiodic plasmonic array platforms,'' {\em MRS Commun.}, vol.~6, no.~2, pp.~105--115, 2016.

\bibitem{2013-Stolk-JoCP}
C.~C. Stolk, ``A rapidly converging domain decomposition method for the {H}elmholtz equation,'' {\em J. Comput. Phys.}, vol.~241, pp.~240--252, 2013.

\bibitem{2019-Gander-SR}
M.~J. Gander and H.~Zhang, ``A class of iterative solvers for the {H}elmholtz equation: Factorizations, sweeping preconditioners, source transfer, single layer potentials, polarized traces, and optimized {S}chwarz methods,'' {\em SIAM Rev.}, vol.~61, no.~1, pp.~3--76, 2019.

\bibitem{2019-Leng-SJoSC}
W.~Leng and L.~Ju, ``An additive overlapping domain decomposition method for the {H}elmholtz equation,'' {\em SIAM J. Sci. Comput.}, vol.~41, no.~2, pp.~A1252--A1277, 2019.

\bibitem{2024-Gao-ITAP}
H.-W. Gao, X.-M. Xin, Q.~J. Lim, S.~Wang, and Z.~Peng, ``Efficient full-wave simulation of large-scale metasurfaces and metamaterials,'' {\em IEEE Trans. Antennas Propag.}, vol.~72, no.~1, pp.~800--811, 2024.

\bibitem{2019-Liu-IToMTaT}
J.~Liu, G.~Cai, J.~Yao, N.~Liu, and Q.~H. Liu, ``Spectral numerical mode matching method for metasurfaces,'' {\em IEEE Trans. Microwave Theory Tech.}, vol.~67, no.~7, pp.~2629--2639, 2019.

\bibitem{2020-Liu-ITMTT}
J.~Liu, N.~Liu, and Q.~H. Liu, ``Microscopic modeling of metasurfaces by the mixed finite element numerical mode-matching method,'' {\em IEEE Trans. Microwave Theory Tech.}, vol.~68, no.~2, pp.~469--478, 2020.

\bibitem{2024-Zhang-PRE}
N.~Zhang and Y.~Y. Lu, ``Spectral {G}alerkin mode-matching method for applications in photonics,'' {\em Phys. Rev. E}, vol.~109, no.~5, p.~055303, 2024.

\bibitem{2020-Christiansen-OE}
R.~E. Christiansen, Z.~Lin, C.~Roques-Carmes, Y.~Salamin, S.~E. Kooi, J.~D. Joannopoulos, M.~Soljačić, and S.~G. Johnson, ``Fullwave {M}axwell inverse design of axisymmetric, tunable, and multi-scale multi-wavelength metalenses,'' {\em Opt. Express}, vol.~28, no.~23, p.~33854, 2020.

\bibitem{2023-Xue-apa}
W.~Xue, H.~Zhang, A.~Gopal, V.~Rokhlin, and O.~D. Miller, ``Fullwave design of cm-scale cylindrical metasurfaces via fast direct solvers,'' {\em arXiv preprint arXiv:2308.08569}, 2023.

\bibitem{2010-Oskooi-CPC}
A.~F. Oskooi, D.~Roundy, M.~Ibanescu, P.~Bermel, J.~Joannopoulos, and S.~G. Johnson, ``{MEEP}: A flexible free-software package for electromagnetic simulations by the fdtd method,'' {\em Comput. Phys. Comm.}, vol.~181, no.~3, pp.~687--702, 2010.

\bibitem{2019-Warren-CPC}
C.~Warren, A.~Giannopoulos, A.~Gray, I.~Giannakis, A.~Patterson, L.~Wetter, and A.~Hamrah, ``A {CUDA}-based {GPU} engine for {gprMax}: Open source {FDTD} electromagnetic simulation software,'' {\em Comput. Phys. Comm.}, vol.~237, pp.~208--218, 2019.

\bibitem{2021-Hughes-APL}
T.~W. Hughes, M.~Minkov, V.~Liu, Z.~Yu, and S.~Fan, ``A perspective on the pathway toward full wave simulation of large area metalenses,'' {\em Appl. Phys. Lett.}, vol.~119, no.~15, 2021.

\bibitem{2016-Byrnes-OE}
S.~J. Byrnes, A.~Lenef, F.~Aieta, and F.~Capasso, ``Designing large, high-efficiency, high-numerical-aperture, transmissive meta-lenses for visible light,'' {\em Opt. Express}, vol.~24, no.~5, p.~5110, 2016.

\bibitem{2018-Chen-NN}
W.~T. Chen, A.~Y. Zhu, V.~Sanjeev, M.~Khorasaninejad, Z.~Shi, E.~Lee, and F.~Capasso, ``A broadband achromatic metalens for focusing and imaging in the visible,'' {\em Nat. Nanotechnol.}, vol.~13, no.~3, pp.~220--226, 2018.

\bibitem{2018-Pestourie-OE}
R.~Pestourie, C.~Pérez-Arancibia, Z.~Lin, W.~Shin, F.~Capasso, and S.~G. Johnson, ``Inverse design of large-area metasurfaces,'' {\em Opt. Express}, vol.~26, no.~26, p.~33732, 2018.

\bibitem{2019-Lin-OE}
Z.~Lin and S.~G. Johnson, ``Overlapping domains for topology optimization of large-area metasurfaces,'' {\em Opt. Express}, vol.~27, no.~22, p.~32445, 2019.

\bibitem{2008-Hu-OE}
Z.~Hu and Y.~Y. Lu, ``Efficient analysis of photonic crystal devices by {D}irichlet-to-{N}eumann maps,'' {\em Opt. Express}, vol.~16, no.~22, p.~17383, 2008.

\bibitem{2012-Lu-JCP}
W.~Lu and Y.~Y. Lu, ``Waveguide mode solver based on {N}eumann-to-{D}irichlet operators and boundary integral equations,'' {\em J. Comput. Phys.}, vol.~231, no.~4, pp.~1360--1371, 2012.

\bibitem{1996-Berenger-JoCP}
J.-P. Berenger, ``Three-dimensional perfectly matched layer for the absorption of electromagnetic waves,'' {\em J. Comput. Phys.}, vol.~127, no.~2, pp.~363--379, 1996.

\bibitem{1996-Gedney-IToAaP}
S.~D. Gedney, ``An anisotropic perfectly matched layer-absorbing medium for the truncation of {FDTD} lattices,'' {\em IEEE Trans. Antennas Propag.}, vol.~44, no.~12, pp.~1630--1639, 1996.

\bibitem{2024-Jiang-AMC}
X.~Jiang, Z.~Sun, L.~Sun, and Q.~Ma, ``An adaptive finite element {PML} method for {Helmholtz} equations in periodic heterogeneous media,'' {\em Appl. Math. Comput.}, vol.~43, no.~4, 2024.

\bibitem{2020-Lyche-Numerical}
T.~Lyche, {\em Numerical linear algebra and matrix factorizations}.
\newblock Springer International Publishing, 2020.

\bibitem{2014-Lin-S}
D.~Lin, P.~Fan, E.~Hasman, and M.~L. Brongersma, ``Dielectric gradient metasurface optical elements,'' {\em Science}, vol.~345, no.~6194, pp.~298--302, 2014.

\bibitem{2020-Lawrence-NN}
M.~Lawrence, D.~R. Barton, J.~Dixon, J.-H. Song, J.~van~de Groep, M.~L. Brongersma, and J.~A. Dionne, ``High quality factor phase gradient metasurfaces,'' {\em Nat. Nanotechnol.}, vol.~15, no.~11, pp.~956--961, 2020.

\bibitem{2024-Polyanskiy-Refractiveindex.info}
M.~N. Polyanskiy, ``Refractiveindex.info database of optical constants,'' {\em Scientific Data}, vol.~11, no.~1, 2024.

\end{thebibliography}

\appendix
  \section{Additional steps for boundary relationships}\label{APP:Omegaj}
  At the end of Sec. 3, we only described the two steps for constructing the NtD
  operator $\NtD{j}$ and the inhomogeneous vector $\RTM{j}$ in Eq.
  \eqref{NtD:Omegaj}. Here, we discuss the detail for obtaining Eq.
  \eqref{NtD:Omegaj} from Eqs. \eqref{NtD:j1} and \eqref{NtD:j02}. Using Eq.
  \eqref{waveu:wj}, the equations with right-hand side terms $\cmdw{j}{1,-}$ and
  $\cmdw{j}{2,+}$ in Eqs. \eqref{NtD:j02} can be reformulated as
  \begin{eqnarray}\label{NtD:j0:eqn3}
    \cmdw{j}{1,+}-\cmdwi{j}{1}=
    \NtD{j,0}{31}\cmdpxu{j}{0}+\NtD{j,0}{32}\cmdpxu{j+1}{0} +\NtD{j,0}{33}(\cmdpyw{j}{1,+}-\cmdpywi{j}{1}),\\
    \cmdw{j}{2,-}-\cmdwi{j}{2}=\NtD{j,2}{11}(\cmdpyw{j}{2,-}-\cmdpywi{j}{2})+ \NtD{j,2}{12}\cmdpxu{j}{2}+\NtD{j,2}{13}\cmdpxu{j+1}{2}.\label{NtD:j2:eqn1}
  \end{eqnarray}

  Substitute the first and fourth members of Eqs. \eqref{NtD:j1} into Eqs.
  \eqref{NtD:j0:eqn3} and \eqref{NtD:j2:eqn1}, respectively, we have
  \begin{eqnarray}\label{NtD:PyU_cellj}
    \Lambda_{w}\begin{bmatrix}
      \cmdpyw{j}{2,-} \\
      \cmdpyw{j}{1,+} \\
    \end{bmatrix}=V_1\begin{bmatrix}
      \cmdpxu{j}   \\
      \cmdpxu{j+1} \\
    \end{bmatrix}+
    \begin{bmatrix}
      \cmdwi{j}{2} \\
      \cmdwi{j}{1} \\
    \end{bmatrix}-\begin{bmatrix}
      \NtD{j,2}{11} &               \\
                    & \NtD{j,0}{33} \\
    \end{bmatrix}
    \begin{bmatrix}
      \cmdpywi{j}{2} \\
      \cmdpywi{j}{1} \\
    \end{bmatrix},
  \end{eqnarray}
  where
  \begin{eqnarray*}
    \Lambda_{w}= \begin{bmatrix}
      \NtD{j,1}{44}-\NtD{j,2}{11} & \NtD{j,1}{41}               \\
      \NtD{j,1}{14}               & \NtD{j,1}{11}-\NtD{j,0}{33} \\
    \end{bmatrix},u_{j}=\begin{bmatrix}
      \cmdu{j}{0} \\\cmdu{j}{1}\\ \cmdu{j}{2}\\
    \end{bmatrix},\\
    V_1= \begin{bmatrix}
      0             & -\NtD{j,1}{42} & \NtD{j,2}{12} & 0             & -\NtD{j,1}{43} & \NtD{j,2}{13} \\
      \NtD{j,0}{31} & -\NtD{j,1}{12} & 0             & \NtD{j,0}{32} & -\NtD{j,1}{13} & 0             \\
    \end{bmatrix}.
  \end{eqnarray*}

  Let $\Lambda_{w}^{-1}$ be the inverse operator of $\Lambda_{w}$, we can
  solve Eq. \eqref{NtD:PyU_cellj} for $\cmdpyw{j}{2,-}$ and $\cmdpyw{j}{1,+}$ as
  follows:
  \begin{eqnarray}\label{Eqn:PyU}
    \begin{bmatrix}
      \cmdpyw{j}{2,-} \\
      \cmdpyw{j}{1,+} \\
    \end{bmatrix}=\Lambda_{w}^{-1}V_1\begin{bmatrix}
      \cmdpxu{j}   \\
      \cmdpxu{j+1} \\
    \end{bmatrix}+\RTM{j}{w},
  \end{eqnarray}
  where 
  \begin{eqnarray*}
    \RTM{j}{w}=\Lambda_{w}^{-1}
    \begin{bmatrix}
      \cmdwi{j}{2} \\
      \cmdwi{j}{1} \\
    \end{bmatrix}-\Lambda_{w}^{-1}
    \begin{bmatrix}
      \NtD{j,2}{11} &               \\
                    & \NtD{j,0}{33} \\
    \end{bmatrix}
    \begin{bmatrix}
      \cmdpywi{j}{2} \\
      \cmdpywi{j}{1} \\
    \end{bmatrix}.
  \end{eqnarray*}

  The equations with right-hand-side terms $\cmdu{j}{k}$ and $\cmdu{j+1}{k}$
  ($j=0,1,2$) in Eqs. \eqref{NtD:j1} and \eqref{NtD:j02} can be written as
  \begin{eqnarray}\label{NtD:Cellj_0}
    \begin{bmatrix}
      u_{j}   \\
      u_{j+1} \\
    \end{bmatrix}=G^{(j)}\begin{bmatrix}
      \cmdpxu{j}   \\
      \cmdpxu{j+1} \\
    \end{bmatrix}+V_3\begin{bmatrix}
      \cmdpyw{j}{2,-} \\
      \cmdpyw{j}{1,+} \\
    \end{bmatrix}-V_4\begin{bmatrix}
      \cmdpywi{j}{2} \\
      \cmdpywi{j}{1} \\
    \end{bmatrix},
  \end{eqnarray}
  where
  {\small
  \begin{eqnarray*}
    G^{(j)}=\begin{bmatrix}
      \NtD{j,0}{11} &               &               & \NtD{j,0}{12} &               &               \\
                    & \NtD{j,1}{22} &               &               & \NtD{j,1}{23} &               \\
                    &               & \NtD{j,2}{22} &               &               & \NtD{j,2}{23} \\
      \NtD{j,0}{21} &               &               & \NtD{j,0}{22} &               &               \\
                    & \NtD{j,1}{32} &               &               & \NtD{j,1}{33} &               \\
                    &               & \NtD{j,2}{32} &               &               & \NtD{j,2}{33} \\
    \end{bmatrix},~~V_3=
    \begin{bmatrix}
                    & \NtD{j,0}{13} \\
      \NtD{j,1}{24} & \NtD{j,1}{21} \\
      \NtD{j,2}{21} &               \\
                    & \NtD{j,0}{23} \\
      \NtD{j,1}{34} & \NtD{j,1}{31} \\
      \NtD{j,2}{31} &               \\
    \end{bmatrix},~~V_4=
    \begin{bmatrix}
                    & \NtD{j,0}{13} \\
                    &               \\
      \NtD{j,2}{21} &               \\
                    & \NtD{j,0}{23} \\
                    &               \\
      \NtD{j,2}{31} &               \\
    \end{bmatrix}.
  \end{eqnarray*}
  }
  Substitute Eq. \eqref{Eqn:PyU} into Eq. \eqref{NtD:Cellj_0}, we have
  \begin{eqnarray}\label{NtD:j}
    \begin{bmatrix}
      u_{j}   \\
      u_{j+1} \\
    \end{bmatrix}=\Lambda^{(j)}\begin{bmatrix}
      \cmdpxu{j}   \\
      \cmdpxu{j+1} \\
    \end{bmatrix}+
    \RTM{j}{},~
    \Lambda^{(j)}=G^{(j)}+V_3A_{w}^{-1}V_1,~
    \RTM{j}{}=V_3 \RTM{j}{w}-V_4\begin{bmatrix}
      \cmdpywi{j}{2} \\
      \cmdpywi{j}{1} \\
    \end{bmatrix}.
  \end{eqnarray}

  \section{Approximation by local function expansion}\label{Approx:vectors}
  In Sec. \ref{Sec:approximation}, to make the content simple, we simplify the
  notations of some quantities, such as NtD matrices, approximations for some
  functions and the related column vectors. Here, we give the following
  approximations for all functions
  {\small
  \begin{eqnarray}\label{approx:wj}
    \cmdw{j}{k,\pm}\approx\lsum{m=1}{M_{j}}\cmdwh{j}{k,\pm}{m}\phi_{j,m}(x),&
    \cmdpyw{j}{k,\pm}\approx\lsum{m=1}{M_{j}}\cmdpywh{j}{k,\pm}{m}\phi_{j,m}(x),&x\in(x_{j},x_{j+1}),~k=1,2,\\
    \cmdu{j}{k}\approx\lsum{m=1}{N_k}\cmduh{j}{k}{m}\varphi_{k,m}(y),&
    \cmdpxu{j}{k}\approx\lsum{m=1}{N_k}\cmdpxuh{j}{k}{m}\varphi_{k,m}(y),&y\in(y_{k},y_{k+1}),~k=0,1,2,\label{approx:uj}\\
    \cmdwi{j}{k}\approx \lsum{m=1}{M_{j}}\cmdwih{j}{k}{m}\phi_{j,m}(x), &
    \cmdpywi{j}{k}\approx \lsum{m=1}{M_{j}}\cmdpywih{j}{k}{m}\phi_{j,m}(x),&x\in(x_{j},x_{j+1}),~k=1,2,\label{approx:wij}\\
    \cmdui{1}{1}\approx\lsum{m=1}{N_1}\cmduih{1}{1}{m}\varphi_{1,m}(y),&
    \cmdpxui{1}{1}\approx\lsum{m=1}{N_1}\cmdpxuih{1}{1}{m}\varphi_{1,m}(y),& y\in(y_{1},y_{2}).\label{approx:uij}
  \end{eqnarray}
  }

  In Eqs. \eqref{NTDM:j012} and \eqref{NtDM:0andT}, there are some column vectors
  (e.g. $\cmduh{j}{1}$, $\cmdpxuh{j}{1}$), their elements are the expansion
  coefficients of the corresponding approximations (Eqs. \eqref{approx:wj} -
  \eqref{approx:uij}), we add a subscript $m$ to a vector to denote its $m$th
  element. These column vectors are
  {\small
  \begin{eqnarray*}
    \cmduh{j}{k}=\left[
      \cmduh{j}{k}{1} ,
      \cdots ,
      \cmduh{j}{k}{N_{k}}
      \right],&
    \cmdpxuh{j}{k}=\left[\cmdpxuh{j}{k}{1},
      \cdots ,
      \cmdpxuh{j}{k}{N_{k}}\right],&1\leq j\leq\NTC,~k=0,1,2,\\
    \cmdwh{j}{k,\pm}=\left[
      \cmdwh{j}{k,\pm}{1} ,
      \cdots ,
      \cmdwh{j}{k,\pm}{M_{j}}
      \right],&
    \cmdpywh{j}{k,\pm}=\left[
      \cmdpywh{j}{k,\pm}{1},
      \cdots ,
      \cmdpywh{j}{k,\pm}{M_{j}}
      \right],&1\leq j<\NTC,~k=1,2,\\
    \cmduih{j}{1}=\left[
      \cmduih{j}{1}{1} ,
      \cdots,
      \cmduih{j}{1}{N_1}
      \right],&
    \cmdpxuih{j}{1}=\left[
      \cmdpxuih{j}{1}{1} ,
      \cdots ,
      \cmdpxuih{j}{1}{N_1}
      \right],&j=1, \NTC,\\
    \cmdwih{j}{k}=\left[
      \cmdwih{j}{k}{1} ,
      \cdots   ,
      \cmdwih{j}{k}{M_{j}} ,
      \right],&
    \cmdpywih{j}{k}=\left[
      \cmdpywih{j}{k}{1} ,
      \cdots ,
      \cmdpywih{j}{k}{M_{j}}
      \right],&1\leq j<\NTC.
  \end{eqnarray*}
  }

\end{document}